\documentclass[aps,prl,reprint,superscriptaddress,amsmath, amssymb]{revtex4-2}

\usepackage{graphicx}% Include figure files
\usepackage{dcolumn}% Align table columns on decimal point
\usepackage{bm}% bold math
\usepackage{graphicx}
\usepackage{siunitx}
\usepackage{color}
\usepackage[dvipsnames]{xcolor}
\usepackage{physics}

\DeclareMathOperator{\e}{e}
\newcommand{\s}{\mathrm{s}}
\newcommand{\p}{\mathrm{p}}

\begin{document}

%\preprint{}

\title{Resonant propagation of x-rays from the linear to the nonlinear regime}


\author{Kai Li}
%\thanks{}
\affiliation{Department of Physics, The University of Chicago, Chicago, Illinois 60637, USA}
\affiliation{Chemical Sciences and Engineering Division, Argonne National Laboratory, Argonne, Illinois 60439, USA}
\author{Marie Labeye}
\affiliation{Department of Physics and Astronomy, Louisiana State University, Baton Rouge, LA 70803-4001, USA}
\affiliation{PASTEUR, D\'epartement de Chimie, \'Ecole Normale Sup\'erieure, PSL University, Sorbonne Universit\'e, CNRS, 75005 Paris, France}
\author{Phay J. Ho}
\affiliation{Chemical Sciences and Engineering Division, Argonne National Laboratory, Argonne, Illinois 60439, USA}

\author{Mette B. Gaarde}
\email{mgaarde1@lsu.edu}
\affiliation{Department of Physics and Astronomy, Louisiana State University, Baton Rouge, LA 70803-4001, USA}

\author{Linda Young}
\email{young@anl.gov}
\affiliation{Chemical Sciences and Engineering Division, Argonne National Laboratory, Argonne, Illinois 60439, USA}
\affiliation{Department of Physics and James Franck Institute, The University of Chicago, Chicago, Illinois 60637, USA}


\date{\today}

\begin{abstract}
 %The modification of strong x-ray fields propagating through a resonant medium of gaseous neon is studied via simulation. The simulation is based upon the solution of a 3D time-dependent Schrödinger-Maxwell equation, with the incident x-ray photon energy on resonance with 1s-3p transition.  We solved for the evolution of the combined incident and medium-generated fields – which includes stimulated emission, absorption, ionization and Auger decay, as a function of the input pulse energy and duration. Stimulated Raman scattering of core-excited states to $2p^{-1}3p$ occurs at high x-ray intensity. The nonlinear interaction also induces pulse reshaping and spectral modulation which is of interest for x-ray optics and spectroscopy. Moreover, self-induced transparency (SIT) and self-focusing of strong x-ray free-electron laser (XFEL) pulses were explored. The differences between two level and three level system reveal a new mechanism of SIT and self-focusing associated with stimulated Raman scattering. These effects are important to understand and potentially applicable as control variables for XFEL pulse properties.
 
We present a theoretical study of temporal, spectral, and spatial reshaping of intense, ultrafast x-ray pulses propagating through a resonant medium. Our calculations are based on the solution of a 3D time-dependent Schr\"odinger-Maxwell equation, with the incident x-ray photon energy on resonance with the core-level 1s-3p transition in neon. We study the evolution of the combined incident and medium-generated field, including the effects of stimulated emission, absorption, ionization and Auger decay, as a function of the input pulse energy and duration. We find that stimulated Raman scattering between core-excited states $1s^{-1}3p$ and $2p^{-1}3p$ occurs at high x-ray intensity, and that the emission around this frequency is strongly enhanced when also including the similar $1s^{-1}-2p^{-1}$ response of the ion. We also explore the dependence of x-ray self-induced transparency (SIT) and self-focusing on the pulse intensity and duration, and we find that the stimulated Raman scattering plays an important role in both effects. Finally, we discuss how these nonlinear effects may potentially be exploited as control parameters for pulse properties of x-ray free-electron laser sources. 
\end{abstract}

%\keywords{}

\maketitle

\section{introduction}
%Understanding the fundamental electromagnetic radiation and matter interactions at high intensity has long been a vibrant scientific frontier. The nonlinear phenomena enabled by strong photon sources in microwave and optical regions has been utilized to control nuclear and valence electronic transitions which leads to breakthroughs in many fields of science, such as medical imaging, telecommunication, and the creation and manipulation of novel materials. Exploiting relativistic electron beam coherent radiation, modern x-ray free electron laser (XFEL) facilities delivering unprecedented high intense x-ray up to $10^{19}\, W/cm^2$ makes it possible to study nonlinear effects in x-ray region \cite{emma2010first,ishikawa2012compact}. Recently, electronic multiple ionization \cite{young2010femtosecond}, x-ray stimulated emission \cite{rohringer2012atomic,beye2013stimulated}, second-harmonic generation \cite{shwartz2014x}, and optical and x-ray wave maxing \cite{glover2012x} have been demonstrated using self-amplified spontaneous emission (SASE) XFELs.

Understanding the fundamental interaction between matter and high-intensity electromagnetic radiation has long been a vibrant scientific frontier. %Matter responds to external electromagnetic radiation through microscopic and macroscopic charge displacement which starts to deviate from a linear response with increasing intensity. 
The nonlinear phenomena enabled by intense light sources in the optical, infrared, and microwave regions have been utilized to control electronic, nuclear, and spin transitions which has enabled breakthroughs across many fields of science, such as medical imaging, telecommunication, and the creation and manipulation of novel materials. So far it has been challenging to investigate coherent nonlinear effects at x-ray wavelengths due to their extremely small cross sections. However, modern x-ray free electron laser (XFEL) facilities delivering x-rays at unprecedented intensities up to \SI{e19}{W/cm^2} now make it possible to also study nonlinear effects in x-ray region \cite{emma2010first,ishikawa2012compact}. Although the majority of x-ray nonlinear interaction studies have involved sequential multiphoton ionization processes in thin targets, leading to normally inaccessible charge states \cite{young2010femtosecond,ho2014theoretical,Rudek2012NatPho,rudenko2017femtosecond,Fukuzawa2012PRL},  two-photon absorption \cite{doumy2011nonlinear,tamasaku2014x,Ghimire2015PRA}, second-harmonic generation \cite{shwartz2014x} and optical-x-ray wavemixing \cite{glover2012x} have also been demonstrated in the x-ray regime. 
%two photon Compton scattering \cite{fuchs2015anomalous}. 
The ability of XFEL pulses to efficiently populate core-excited states, and thus induce a population inversion through a thick target, has enabled the observation of x-ray stimulated emission and x-ray stimulated Raman scattering (SRS) \cite{rohringer2012atomic,beye2013stimulated, weninger2013stimulatedPRL,Yoneda2015Nat}. 

A different aspect of nonlinear laser-matter interaction is manifested in the propagation of an intense, short pulse through a dense medium with a resonance that is long-lived relative to the pulse duration. The long-lived free-induction decay \cite{BlochPR1946,brewerPRA1972} that persists after the pulse leads upon propagation to modifications in both the spectral, temporal, and spatial profile of the short pulse \cite{mccall1969self,mccall1967self,crisp1970propagation,chiao1964self,gibbs1976coherent,wright1973self} such as self-induced transparency (SIT) \cite{mccall1969self, mccall1967self}, Burnham-Chiao oscillations in the temporal profile \cite{Burnham1969PR}, and self focusing \cite{chiao1964self, gibbs1976coherent, wright1973self}, and  these phenomena have been extensively investigated in the optical regime. More recently, reshaping of extreme ultraviolet pulses through collective effects has been demonstrated, using either high-harmonic generation \cite{Pfeiffer2013.MacroscopicAbsorption, liao2015beyond, bengtsson_spacetime_2017} or FEL sources \cite{Harries2018PRL} with synchronized infrared pulses as a means of imposing nonlinear processes. The x-ray regime presents a challenge given that inner-shell resonances typically have lifetimes of just a few femtoseconds, and that intense x-ray pulses based on self-amplified spontaneous emission (SASE) have durations of tens to hundreds of femtoseconds and exhibit poor temporal coherence. However, the recent demonstrations of  sub-femtosecond, nearly transform-limited soft and hard x-ray pulses promise a new regime of coherent ultrafast x-ray matter interactions  \cite{marinelli2017experimental, huang2017generating, duris2020tunable}. %{\bf MG: Do we also want to say something about future experimental capabilities for careful characterization of both initial and final spectrum? Or is that too much detail?}

In this paper, we study the spectral, temporal and spatial reshaping of sub- and few-femtosecond, temporally coherent x-ray pulses as they propagate through an atmospheric density neon gas on resonance with the $1s\rightarrow3p$ transition. We introduce a method of simultaneously calculating the quantum dynamics of the neutral and singly ionized neon atoms, represented by few-level systems, and solving the three-dimensional Maxwell wave equation (MWE) to incorporate the collective response of the nonlinear medium. This approach allows us to treat all the different linear and nonlinear processes on an equal footing at both the microscopic and the macroscopic level. We find that stimulated Raman scattering (SRS) between core-excited states $1s^{-1}3p$ and $2p^{-1}3p$ occurs at high x-ray intensity \cite{sun2010propagation, weninger2013stimulatedPRL}, and that it is strongly enhanced when also including the $1s^{-1}-2p^{-1}$ (x-ray lasing) response of the ion. We observe and distinguish the growth of the SRS channel in the neutral relative to the stimulated emission in the ion \cite{rohringer2012atomic,weninger2013PRA}. We also find that at high, but reachable, intensities above \SI{e19}{W/cm^2} a transmitted spectrum with unexpected spectral strength  between the SRS and resonant peaks develops during propagation, and that this can be understood as a generalized Rabi-cycling in an intense multi-color field consisting of both initial and generated frequency components in the pulse. Finally, we investigate the spatial reshaping of the x-ray pulse and find that it is influenced both by SIT and by true self-focusing, and that this can be interpreted in terms of the intensity dependent, and therefore radially varying, modification of the amplitude and phase of the polarization field by the strong x-ray pulse. Finally, we show that all of the effects mentioned above still remain if we substitute the temporally coherent x-ray pulse with a SASE pulse. 

This paper is organized as follows.  Section II describes our theoretical approach to understanding linear and non-linear effects in x-ray propagation through a thick resonant medium.  Section III discusses a number of results for propagation through a neon gas with radiation resonant with the strong core-level $1s\rightarrow3p$ transition: (A) effects of including various ionization continua on the transmitted spectral and temporal profiles, (B) spectral and temporal profiles as a function of intensity and pulse duration, and (C) self induced transparency and self-focussing as a function. In section III(D) we generalize the results in the previous sections obtained for coherent pulses to resonant propagation of SASE pulses. Finally in Section IV, we summarize our findings.

\section{theoretical approach\label{sec:theory}}

In order to model the spatiotemporal reshaping of an  x-ray pulse due to absorption and emission processes we have developed a versatile code based on coupling the 3D MWE to a TDSE solution for a few-level system that can describe both the neutral and ionized species of the system, extending previous x-ray propagation calculations in one spatial dimension \cite{sun2010propagation, weninger2013PRA, weninger2013stimulatedPRL, lyu2020narrow, Harries2018PRL}. For the TDSE solution, we use a hybrid approach in which the evolution of the states in the neutral is treated using a wave-function-based formalism, whereas states in the ionized species are treated using a density matrix formalism. For the neutral, this includes coherent dipole couplings between relevant states, incoherent loss terms due to Auger decay of core-excited states, and incoherent loss terms due to direct ionization out of the relevant core and/or valence orbitals. For the ionized species, the {\it population} of each state is directly related to the ionization out of the equivalent orbital in the neutral, where the relationship is enforced via energy conservation in the field+gas system (see more details below), and the {\it coherences} in the ion are then allowed to develop in the field as it coherently couples different ionic states to each other. In the next couple of paragraphs, we outline this approach in more detail.

\begin{figure}[t]
\centering
	\includegraphics[width=\linewidth]{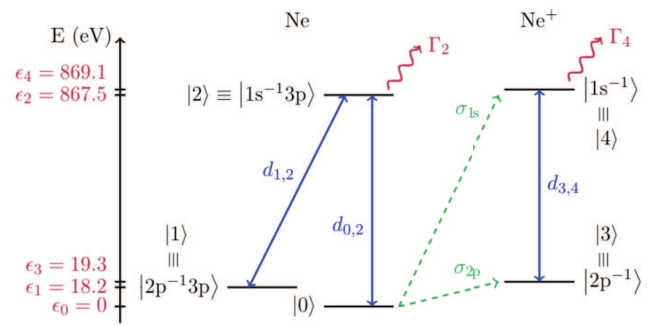}
	\caption{\label{Fig:NeLevels} Schematic of the  energy levels, dipole couplings, and decay rates in the neon atom (left) and the neon ion (right). The dipole couplings $d_{0,2}=0.0077$ a.u., $d_{1,2}=0.048$ a.u.,$d_{3,4}=0.083$ a.u. are taken from \cite{weninger2013PRA}}
	\label{energy diagram}
\end{figure}

In this paper, we consider a 5-level neon system consisting of three states in the neutral and two in the ion, as illustrated in Fig.~\ref{Fig:NeLevels}. These are the ground state, the $1s^{-1}3p$ core excited and the $2p^{-1}3p$ valence excited states in the neutral, at energies of 867.5 eV and 18.2 eV above the ground state, respectively, and the $2p^{-1}$ and $1s^{-1}$ ground and core excited state of the ion, at energies of 19.3 eV and 869.1 eV above the neutral ground state. The coherent and incoherent couplings between the states are illustrated in Fig.~\ref{Fig:NeLevels}, including the magnitudes of the dipole couplings, in a.u. The Auger decay rates of the two core excited states have values $\Gamma_2=\Gamma_4 = 0.0099$ a.u., and the ionization cross sections are $\sigma_{2p} = 0.0084$~Mbarn and $\sigma_{1s} = 0.30$~Mbarn.  The transition dipoles $d_{0,2}, d_{1,2}, d_{3,4}$ from Ref. \cite{weninger2013PRA} were originally calculated with \cite{Cowan} and correspond to a peak absorption cross section for of 1.98 Mb for the $1s\rightarrow1s^{-1}3p$ neutral Ne transition.  %This is in  reasonable agreement with the more recent experimental and theoretical determination of 2.6 Mb \cite{Muller2017ApJ}. %{\bf LY: These photoionization cross sections are perhaps for the continuum but are not accurate. A recent paper by Muller Astrophysical Journal (2017) gives both the Ne and the Ne+ cross sections accurately.  We should probably use those.} {\it MG: has this comment been dealt with?}
We do not include spontaneous emission as a means of initiating the SRS transitions; the spectral background provided by the combination of the initial pulse and the 2.4 fs exponential decay of the resonant polarization is sufficient to provide a seed, see also discussions in \cite{sun2010propagation, weninger2013stimulatedPRL}. 

Our hybrid approach means we are solving a set of six coupled differential equations for the three wave function coefficients $\Psi_0, \Psi_1, \Psi_2$, the two populations $\rho_{3,3}$ and $\rho_{4,4}$, and the coherence between the ionic states $\rho_{3,4}$, in atomic units:

\begin{subequations}
\label{eq:TDSE}
\begin{align}
    \begin{split}
    i\dot{\Psi}_0(t)=& \textcolor{purple}{-i\left(\frac{\Gamma_{1\s}(t)}{2}+\frac{\Gamma_{2\p}(t)}{2}\right)\Psi_0(t)} \\
  &  + \textcolor{blue}{d_{0,2}\mathcal{E}(t) \Psi_2(t)} 
    \end{split}\\
  i\dot{\Psi}_1(t)=&\textcolor{purple}{\epsilon_1\Psi_1(t)} + \textcolor{blue}{d_{1,2}\mathcal{E}(t) \Psi_2(t)}\\
  \begin{split}
  i\dot{\Psi}_2(t)=&\textcolor{purple}{\left(\epsilon_2-i\frac{\Gamma_2}{2}\right)\Psi_2(t)} \\
  &+ \textcolor{blue}{d_{0,2}\mathcal{E}(t) \Psi_0(t) + d_{1,2}\mathcal{E}(t) \Psi_1(t)}
  \end{split}\\[0.2cm]
  \dot{\rho}_{3,3}(t)=&\textcolor{Green}{\Gamma_{2\p}(t)|\Psi_0(t)|^2} \textcolor{blue}{-2id_{3,4}\mathcal{E}(t)\Im(\rho_{3,4}(t))\vphantom{\frac{\Gamma_{1\s}}{2}}}\\
  \begin{split}
  \dot{\rho}_{4,4}(t)=&\textcolor{Green}{\Gamma_{1\s}(t)|\Psi_0(t)|^2} \textcolor{purple}{-\Gamma_4\rho_{4,4}(t)} \\
  &+\textcolor{blue}{2id_{3,4}\mathcal{E}(t)\Im(\rho_{3,4}(t))\vphantom{\frac{\Gamma_{1\s}}{2}}}
  \end{split}\\
  \begin{split}
  \dot{\rho}_{3,4}(t)=&\textcolor{purple}{\left[i\left(\epsilon_4-\epsilon_3\right) - \frac{\Gamma_4}{2} \right]\rho_{3,4}(t)} \\
  &\textcolor{blue}{-id_{3,4}\mathcal{E}(t)(\rho_{4,4}(t)-\rho_{3,3}(t))}.
  \end{split}
\end{align}
\end{subequations}

In these equations, the red terms correspond to the phase and amplitude evolution driven by the energies and lifetimes of the different states; the blue terms are the terms driven by field-induced dipole couplings; and the green terms are responsible for incoherently populating the states of the ion via ionization of the neutral ground state. 

%We note that the separation of the time evolution of the neutral and ionic states allows us considerable flexibility in how to treat the time-dependent coherences between pairs of states, which is what drives the absorption and emission of  x-ray radiation. 
First, note that we do not keep track of any coherence between the neutral and the ion, since we do not keep track of the photoelectrons. This is what allows us to treat them separately, each of them in their own formalism, for numerical efficiency. Second, since the only sources of decoherence in the neutral are from photoionization and Auger decays, this can be simply implemented, within the wave function formalism, as loss terms in the TDSE. On the contrary, there are several incoherent processes affecting the ion. The Auger decay from the core hole can also be implemented as a loss term, but the photoionization processes are source terms that increase \emph{incoherently} the population in the ionic states. This cannot be implemented within the wave function formalism, which is why the ion has to be treated in the density matrix formalism. Indeed, the photoionization terms (green terms) will only increase the populations $\rho_{3,3}$ and $\rho_{4,4}$ in the ionic states, \emph{without} affecting the coherence $\rho_{3,4}$. Note that, since the ion dipole is proportional to the coherence, as long as the coherence is zero the ion neither absorbs nor emits light. The coherence in the ion can only build over time through its interaction with the field (blue term). This is the term that will be responsible for absorption or emission of light in the ion. The incoherent population transfer (green term) will not lead to any coherent emission of light. Finally, it is worth noting that the wave function formalism is less computationally demanding, especially for systems with large numbers of states. 
%The former is less computationally intensive, with the split operator algorithm that we use here, it scales as the number of states to the power 2, against a power 3 for the density matrix formalism. 

The 3D MWE is solved in the frequency domain, by space marching through the gas in a frame that moves at the speed of light. The incident field is linearly polarized, and thus the only driven (and phase matched) response from the atoms is linearly polarized as well. In SI units, and with all frequency-dependent quantities also functions of the cylindrical coordinates $r$ and $z$, the MWE along the field polarization takes the form: 
\begin{align}
\begin{split}
\nabla^2_{\perp} \tilde {\cal E}(\omega) + \frac{2i\omega}{c}\frac{\partial \tilde {\cal E}(\omega)}{\partial z}  =  
& - \frac{\omega^2}{\epsilon_0 c^2}\tilde P(\omega)  \\
& -  i\frac{\omega}{c}\rho_{at} \tilde{\sigma}(\omega)\tilde {\cal E}(\omega).
\label{tdse-mwe}
\end{split}
\end{align}
Here $\tilde {\cal E}(\omega)$ is the electric field which contains all the frequencies of the incoming and generated field, and the polarization-field source term $\tilde P(\omega)=\rho_\mathrm{at}\tilde d(\omega)$ is calculated from the single atom dipole moment $\tilde d(\omega)$, including both neutral and ion contribution, via the few-level TDSE solutions described above. The second term on the right-hand side represents the loss of energy from the field due to ionization, where $\tilde{\sigma}(\omega)$ is the effective cross section due to all ionization processes at frequency $\omega$, and $\rho_{at}$ is the density of neutral atoms. 

To ensure energy conservation in the field+gas system this energy loss from the field, which is implemented in a macroscopic sense in the MWE, has to match the energy absorbed at the microscopic level by the atoms in the TDSE through the time dependent ionization rates. The energy absorbed from the field is easily deduced from the cross section with Beer's law. However, we can't know how much energy is absorbed by the atoms before we solve the TDSE. We thus implemented a two step procedure to enforce energy conservation. First we evaluate how many photons each atom is expected to absorb, using the cross section, to compute a time dependent rate that is used to solve the TDSE. Then we extract from the TDSE the exact number of photons absorbed by each atom. Finally we adjust the cross section that we use in the MWE. In the following, we detail how this is implemented. 

For each ionization channel $i$, the frequency dependent ionization cross section $\sigma_i(\omega)$ is taken as an arctangent function centered around the ionization threshold energy $\omega_\mathrm{th}$:
\begin{align}
    \sigma_i(\omega)=\sigma_i\left[ \frac{1}{2}+\arctan(2\frac{\omega-\omega_\mathrm{th}}{\pi\Gamma_\mathrm{ion}}) \right],
\end{align}
where $\sigma_i$ is the ionization cross section, and $\Gamma_\mathrm{ion}$ is the inverse lifetime of the ionic state populated by the ionization process, here it is equal to the Auger decay rate. The arctangent function is the integral of the lorentzian cross sections for each continuum states that can be populated by an $\omega$ photon, under the assumption that the ionized states all have the same lifetimes, and that the cross section is flat \cite{breinig_atomic_1980}. 
%{(\bf ML: for the 2p ionization we used different shape for the cross section. At some point we used a constant ionization cross section over the full spectrum, considering that the ionization threshold was very low compared to the incoming pulse spectrum. I don't know what Kai used in the end, since this is an input parameter.)} {\it MG: should we say why we have the arctan function - to smoothe the abrupt transition? ML: It's actually more physical than that. It's because we integrate over all the continuum states that a photon at energy exactly $\omega$ can populate. For each of these states we have a lorentzian, and the integral of a lorentzian gives an arctangent function.}

The corresponding time-dependent ionization rate $\Gamma_i(t)$ used in the TDSE
%, is computed from the electric field and the frequency dependent ionization cross section $\sigma_i(\omega)$. Its time dependence simply follows the square of 
is computed from this cross section $\sigma_i$ and from the electric field. In usual first order time dependent perturbation theory approaches, the ionization rate does not depend on time, and in the case of a monochromatic incident field of amplitude $F$ the rate is proportional to $F^2$. Here we suppose that we can incoherently sum the contribution from all incident frequencies, and we make the approximation that the rate adiabatically follows the slowly varying field envelope $F(t)$: 
\begin{align}
    &\Gamma_i(t)=\alpha_i F^2(t)\\
    &F(t) = \frac{1}{\pi}\left|\int_0^\infty \tilde{\mathcal{E}}(\omega) \e^{-i\omega t}\mathrm{d}\omega\right|.
\end{align}
To set the proportionality constant $\alpha_i$, we compute the expected number of absorbed photons per atom $N_\mathrm{ph}$ by integrating $\Gamma_i(t)$ over the full pulse. Note that this supposes that the population in the ground state is almost constant. Then, we express $N_\mathrm{ph}$ as the energy $U_i$ lost by the field at a given $z$ position, divided by the number of atoms at $z$ and by the energy of one absorbed photon:
\begin{align}
    &N_\mathrm{ph}=\int_0^{\infty} \Gamma_i(t)\mathrm{d}t= \frac{U_i}{\rho_\mathrm{at}\hbar\omega_i}\\
    &U_i=\frac{\varepsilon_0c}{2\pi\Delta z}\left[\int_{-\infty}^\infty|\tilde{\mathcal{E}}(\omega)|^2\left(1-\e^{-\rho_\mathrm{at}\sigma_i\Delta z}\right)\mathrm{d}\omega\right].
\end{align}
Using the fact that
\begin{align}
\int_{-\infty}^\infty F^2(t)\mathrm{d}t= \frac{1}{\pi}\int_{-\infty}^\infty |\tilde{\mathcal{E}}(\omega)|^2\mathrm{d}\omega
\end{align}
we finally get
\begin{align}
\Gamma_i(t)=\frac{\varepsilon_0c}{2\hbar\omega_i\rho_\mathrm{at}\Delta z}\frac{\int_{-\infty}^\infty|\tilde{\mathcal{E}}(\omega)|^2\left(1-\e^{-\rho_\mathrm{at}\sigma_i\Delta z}\right)\mathrm{d}\omega}{\int_{-\infty}^\infty |\tilde{\mathcal{E}}(\omega)|^2\mathrm{d}\omega}F^2(t)
\end{align}
However, because of all the coherent and incoherent processes taking place simultaneously, the atoms might actually absorb less photons that the expected $N_\mathrm{ph}$, especially if the interaction with the field is nonlinear. To account for this, we keep track of the number of photons $\tilde{N}_\mathrm{ph}$ that are \textit{really} absorbed within each ionization channel $i$ described by the TDSE. This is computed by summing all the population that is transferred from the neutral to the ion during the TDSE propagation. The field+gas energy conservation is enforced by rescaling the ionization cross section $\tilde{\sigma}_i$ that is used in the MWE:
\begin{align}
    \tilde{\sigma}_i(\omega)=\sigma_i(\omega)\frac{\tilde{N}_\mathrm{ph}}{N_\mathrm{ph}}=\sigma_i(\omega)\frac{\tilde{N}_\mathrm{ph}\rho_\mathrm{at}\hbar\omega_i}{U_i},
\end{align}
so that the number of photons macroscopically removed from the field by the MWE exactly corresponds to the number $\tilde{N}_\mathrm{ph}$ of photons that the atoms actually absorbed.
%This ensures that the energy absorbed by the atoms is exactly equal to the energy lost by the field as it propagates through the gaz.
Finally, the effective cross section $\tilde{\sigma}(\omega)$ in the MWE (Eq.~\eqref{tdse-mwe}) is the sum of all the rescaled $\tilde{\sigma}_i(\omega)$ of all the different ionization channels.  This procedure reproduces the experimental photoabsorption cross section with the $1s$-continuum deduced by computing the quantity $\sigma_{\mathrm{abs}} =     $ ln$[I_{\mathrm{incident}}(\omega)/I_{\mathrm{trans}}({\omega})]/nL$, where $n$ and $L$ are the number density and pathlength of the target, respectively. 
%{\bf LY: What is the effective cross section for the 1s ionization in Mb compared to the experimental value?} 

The initial electric field is a focused Gaussian in space, and in the frequency domain is defined as the Fourier transform of the initial pulse. At each plane in the propagation direction, the time-dependent dipole moment is calculated from both the neutral and the ionic systems, $d(t) = d_n(t) + d_i(t)$, and the macroscopic polarization field is proportional to the dipole moment via the density (note that this is the initial density of the gas; the time-dependence of the neutral and ion density is already incorporated into the time dependent wave functions and coherences in \eqref{eq:TDSE} above). Finally, we use the calculated source terms on the right-hand side of Eq.~\ref{tdse-mwe} to propagate the electric field to the next plane in the propagation direction.  This self-consistent approach ensures that both the linear and non-linear response of the atoms to the field is incorporated back into the propagating XUV (and IR) field, and we can therefore treat both linear and non-linear absorption and emission on an equal footing. Note that in contrast to this polarization-field-driven coherent exchange of energy between the field and the gas, the ionization loss term is incoherent. 

The TDSE \eqref{eq:TDSE} is solved with a split-operator algorithm, where the propagation of the neutral wave function on the one hand and of the density matrix of the ion on the other hand are done in parallel. The MWE \eqref{tdse-mwe} is solved with a Crank-Nicolson algorithm. All Fourier transforms are computed with the FFTW package \cite{fftw_article}, and diagonalizations were performed using the LAPACK library \cite{lapack_website}.

%*******************************************************************

\section{discussion of results\label{sec:results}}
\subsection{Role of Ionization Continua\label{sec:results logitudinal}}
\begin{figure*}[!htbp!]
\centering
\setlength{\unitlength}{1\textwidth}
\begin{picture}(1,0.69)
\put(0,0){\includegraphics[width=0.99\textwidth, height=0.5\textheight]{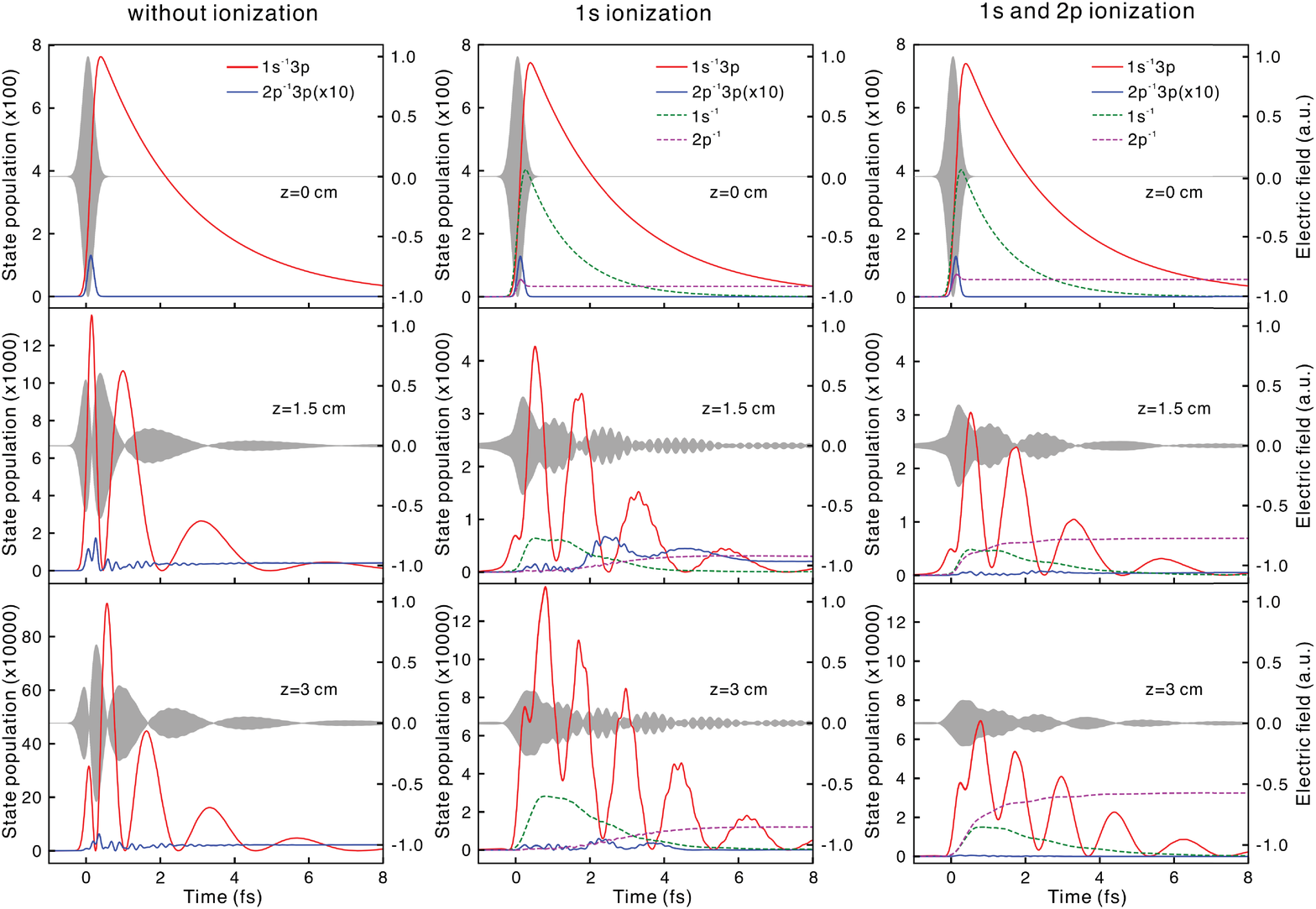}}
\end{picture}
\caption{\label{fig:population} The evolution of the electric field and state populations during resonant propagation  in neon gas of a 0.25-fs (FWHM) x-ray pulse centered on the $1s\rightarrow3p$ Ne transition at 867.5 eV with peak intensity \SI{e18}{W/cm^2}. Time dependent populations of the core-excited state $1s^{-1}3p$, valence-excited state $2p^{-1}3p$ for neutral Ne and 1s and 2p hole states for Ne$^+$ are shown for propagation distances z=0, 1.5 and 3.0 cm. The profile of the electric field at each propagation distance is indicated in gray. The three columns represent the evolution without ionization channels (left), including 1s ionization (center) and including 1s and 2p ionization (right), \\
}
\end{figure*}

\begin{figure*}
\centering
\includegraphics[width=0.99\textwidth, height=0.18\textheight]{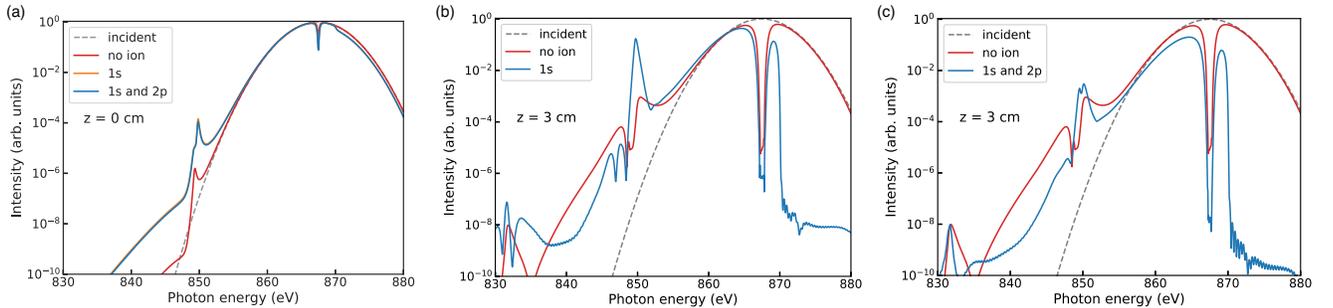}
\caption{\label{fig:3spectrum} Transmitted spectra through Ne gas with different ionization conditions. Incident pulse is 0.25 fs (FWHM) Gaussian, \SI{e18}{W/cm^2}, centered at 867.5 eV. (a)  Spectra after propagating one step (0.05 cm) starting at z=0 for no ionization (red), $1s$ ionization (yellow), $1s+2p$ ionization (blue). (b) Spectra at z=3.0 cm for no ionization and $1s$ ionization. (c) Spectra at z=3.0 cm for no ionization and $1s+2p$ ionization.
}
\end{figure*}

%\begin{figure} [!htbp!]
%\centering
%\setlength{\unitlength}{1\textwidth}
%\includegraphics[width=0.45\textwidth, %height=0.23\textheight]{1e18_spectrum_differentIon_0.05cm.eps}
%\caption{\label{fig:3spectrum}Spectrum of different ionization for propagation distance %z=0 cm. Incident pulse is 0.25 fs (FWHM) Gaussian with \SI{e18}{W/cm^2} peak intensity. %Calculated for the 3-level Ne, without ionization channels (red), including 1s %ionization (yellow) and including 1s and 2p ionization (blue).  
%}
%\end{figure}

We now use this theoretical machinery to explore resonant propagation in neon gas as a function of pulse intensity and duration. First, to give insight into the physics at play we examine the role of ionization in the combined atom-ion-EM-field system.  The effects of non-resonant ionization can be negligible in some cases \cite{sun2010propagation}, but here, where we access the deep 1s inner shell with ultrashort, broad bandwidth pulses, this is not the case.  We specifically explore the effects of ionization by computing three scenarios: no ionization, ionization of the 1s shell and ionization of both 1s and 2p shells which directly provide population in the ion.  

We consider a 0.25-fs (FWHM) Gaussian pulse centered on the $1s\rightarrow3p$ resonance at 867.5 eV with a peak intensity of $10^{18}$ W/cm$^2$ (pulse energy $\sim100\,\mu$J)  %the precise value is $113\,\mu$J%) 
propagating through 3-cm of gaseous neon at atmospheric pressure.  The beam is focused to the center of the cell with a confocal parameter of $b=12.35$ cm corresponding to a focus of $6.12\,\mu$m (FWHM) at the cell center. 
%($1/\sqrt{e}$ beam electric field waist of $w_0=5.2\;\mu$m)$.  
%By way of context, the resonance has a peak absolute cross section of $\sigma=2.6$ Mb \cite{Muller2017ApJ} such that absorption of a low intensity pulse according to Beer's law $I(L) = I_0 e^{-\sigma nL}$ with $\sigma nL\approx200$, is well in the optically-thick, self-induced-transparency regime \cite{mccall1967self,slusher1972PRA}.
For context, the resonance has a peak absorption cross section of $\sigma=1.98$ Mb, such that Beer's law $I(L) = I_0 e^{-\sigma nL}$ with $\sigma nL\approx150$ is well in the optically-thick, self-induced-transparency regime \cite{mccall1967self,slusher1972PRA}. (A more recent high accuracy determination of the resonance cross section is 30\% higher \cite{Muller2017ApJ}, but our observed trends and results are expected to be unchanged.)
Here the pulse duration, $\tau_p$, is less than any relaxation or decoherence timescale (set by the Auger decay of 2.4 fs) such that the pulse has a spectral content that is much broader than the absorption line of the medium.  

The simulation uses 30 points per period, $T_p$ = 4.67 attoseconds, and covers a range of 100 fs.  For propagation from $z=0 - 3$ cm, the electric field and state populations in the time domain, and, the spectra in the frequency domain are written as output files for sequential propagation planes spaced by $\Delta z=0.05$ cm. The propagation calculations were performed at a stepsize of $\Delta z/30$.  The 100-fs time window covers the oscillating pulse tails and corresponds to a 0.04-eV grid in the frequency domain.  This fine grid is sufficient to resolve the Ne absorption spectra which have a natural width of 0.27 eV.  

In Fig. \ref{fig:population} and \ref{fig:3spectrum} we show time-domain and frequency domain results, respectively, for selected propagation distances of between 0 and 3 cm, for the three different ionization scenarios described above. The left column in Fig. \ref{fig:population} shows the results when no ionization continua are included.  The amplitude of the electric field is shown in gray and the populations of two excited states in neutral neon, $1s^{-1}3p$ and $2p^{-1}3p$, in red and blue, respectively.  We observe a dramatic temporal reshaping of the pulse as it propagates through the medium. The response at the entrance slab of the medium (z=0) shows a population of the resonantly excited $1s^{-1}3p$ state of 8\% at the end of the pulse, that subsequently decays with the $1s$ hole lifetime of 2.4 fs. The small blip associated with the $2p^{-1}3p$ state is related to the dressing of the states by the electric field, but does not correspond to real population transfer to the $2p^{-1}3p$ state. 

As we continue to propagate into the medium, a ringing is evident as energy is exchanged between the two resonant levels and the propagating field.  The third level, $2p^{-1}3p$, plays little role. This situation is analogous to that shown in Sun et al. \cite{sun2010propagation} where both 2- and 3-level systems without ionization channels show similar ringing behavior for large-area pulses, i.e input $\pi$ pulse.  The results here show that the use of large area pulses ($>\pi$) is not required to achieve the oscillation in the x-ray regime - a feature long appreciated in the optical community \cite{Burnham1969PR,crisp1970propagation}. We also note that the population of the excited $1s^{-1}3p$ state is out of phase with the electric field demonstrating the energy exchange between the medium and field in this simple configuration. The time-domain ringing, and its evolution with propagation distance, can be understood by considering the evolution of the spectrum (see Fig~\ref{fig:3spectrum}(a,b)): As the pulse propagates a hole is eaten from its spectrum, which creates a beat (ringing) in the time domain between the spectral content above and below the resonance frequency. As the hole widens during propagation, the time scale of the ringing shortens with propagation distance, as observed going from $z=1.5- 3.0$ cm in the bottom two panels. (We note that the inclusion of higher Rydberg levels ($3p, 4p, 5p, 6p$) suppresses the aforementioned Burnham-Chiao time-domain ringing but otherwise merely leads to the expected multiple absorption dips in the transmitted spectrum.) 

%{\bf MG suggestion: Maybe we could have a small figure in between the current Fig. 2 and Fig. 3, showing the final (3 cm) spectra for the three different columns in Fig. 2. This would make it easier to follow our point about what the addition of the ionization process does. Right now the only spectrum we show is the one corresponding to the middle column - depending on how important we fee this point is, it would be good to also show the others. This could be a single-panel, one-column figure.}

In the center column of Fig. \ref{fig:population} we show results when the $1s$-ionization continuum and associated ion states are included,  There is a dramatic difference compared to the left column. A $4\%$ population of the $1s^{-1}$ ion state is observed in the first slab. The presence of the $1s^{-1}$ hole produces dipole coupling at the $1s^{-1}\rightarrow2p^{-1}$ transition frequency that is nearly resonant with corresponding transition in the neutral, $1s^{-1}3p\rightarrow2p^{-1}3p$. This can also be observed in the spectrum in Fig.~\ref{fig:3spectrum}(a) as a two-order-of-magnitude enhancement of the 850 eV signal after propagating only one step (0.05 cm). 
After propagation to 1.5 cm a fine temporal beating of the electric field appears that corresponds to $\Delta t = 2\pi/(\omega_{02}-\omega_{12})=0.22$ fs.  The beating persists throughout propagation and is also manifested in the transmitted spectrum, where the stimulated Raman scattering (SRS) peak is prominent (see Fig. \ref{fig:3spectrum}b). For the no-ionization case in the left column, the transmitted SRS peak is much lower, and the fine temporal beating is nearly absent in the temporal profile.

The right column of Fig. \ref{fig:population}, shows propagation when both the $1s$ and $2p$ ionization continua are included.  The addition of the $2p$ continua affects the temporal profile and populations to a lesser degree than the first inclusion of the $1s$ continuum.  The magnitude of the temporal oscillations is slightly suppressed, but the main features shown in the center column are still apparent.  The major difference is the more prominent appearance of the $2p^{-1}$ ion state, which enables absorption at this photon energy, thus suppressing the 850-eV peak and the associated temporal beating. 

\begin{figure*} [!htbp!]
\centering
\includegraphics[width=0.99\textwidth, height=0.36\textheight]{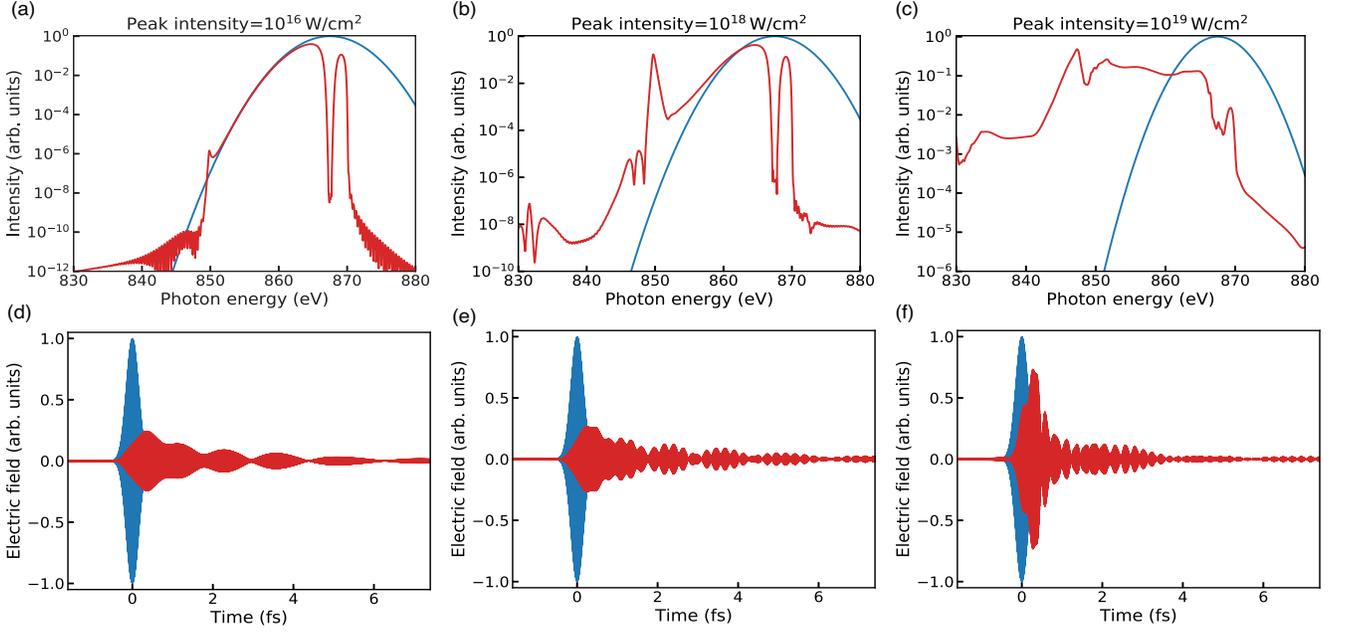}
\caption{\label{fig:intensities}Spectral and temporal profiles for the incident and transmitted pulse at various intensities. Incident pulse is 0.25 fs (FWHM) Gaussian centered at 867.5 eV.  Transmission through 3 cm Ne gas at 1 bar pressure calculated using the five-level Ne, Ne$^+$ and 1s-continuum system. Upper row display spectra for peak intensity (a) \SI{e16}{W/cm^2}, (b) \SI{e18}{W/cm^2} and (c) \SI{e19}{W/cm^2}.  Lower row display temporal profiles:(d) \SI{e16}{W/cm^2}, (e) \SI{e18}{W/cm^2} and (f) \SI{e19}{W/cm^2}  
}
\end{figure*}

% Fig. \ref{fig:3spectrum} illustrates differences in the transmitted spectra for the three ionization conditions with a 0.25-fs, \SI{e18}{W/cm^2} pulse.  The left panel shows the spectra at z=0.  Inclusion of the $1s$ or the $1s+2p$ ionization continua give identical transmitted spectra at $z=0$.  They differ from the spectrum with no ionization, in that both have a feature at $\sim 850$eV 100$\times$ higher than the no-ionization case and both have a new frequency component at 849.8 eV, indicative of the $1s\rightarrow2p$ transition in Ne$^+$.  The middle and right panels show the spectra at $z=3$cm for $1s$ and $1s+2p$ ionizations, respectively. Inclusion of $2p$ ionization enables absorption processes via $1s\rightarrow2p$ transitions in Ne$^+$ and thus decrease the feature at $\sim 850$eV. 

In the following we use as a default the five-level Ne system shown in Fig. \ref{Fig:NeLevels} and the 1s-ionization continuum, i.e. the conditions of the central column in Fig. \ref{fig:population}. This configuration illustrates the main physics involved in the coupled atom-ion-EM field system.  

%***************************************************************

\subsection{Spectral and temporal reshaping as a function of intensity\label{sec:results logitudinal}}

%\begin{itemize}
%    \item resonant absorption, the temporal pulse profile is reshaped, oscillating tails appear, the pulse head is compressed
%    \item stimulated Raman occurs at higher intensities, continuum absorption due to photoionization, fine structure is due to the beats between stimulated Raman and main input pulse (one color to two color)
%    \item with even higher intensity, the spectrum changed a lot, the continuum spectrum appears connecting the stimulated Raman and input pulse (explain as dress atom, stimulated AC stark, one color to milt-color)
%    \item the temporal profile also changes, the oscillating tail was modified and the beats between stimulated Raman and input pulse is still there
%\end{itemize}

%\begin{figure*} [!htbp!]
%\centering
%\setlength{\unitlength}{1\textwidth}
%\begin{picture}(0.9,0.27)
%\put(0,0){\includegraphics[width=0.40\textwidth, height=0.2\textheight]{3levles_0.25fs_1e18_sigma1s_zspec_evolution.eps}}
%\put(0.055,0.23){(a)}
%\put(0.42,0){\includegraphics[width=0.36\textwidth, height=0.2\textheight]{Raman_growth.eps}}
%put(0.485,0.23){(b)}
%\end{picture}
%\caption{\label{fig:SpectralEvolution} Spectral evolution during propagation through the five-level Ne, Ne$^+$ and 1s-continuum system %at 1 bar. (a) Spectral evolution of a 0.25 fs, \SI{e18}{W/cm^2} pulse. (b) Growth of the fraction of transmitted energy contained in the %SRS peak (integrated between 846 - 853 eV) for \SI{e16},\SI{e17}, \SI{e18}{W/cm^2} peak intensities.
%}
%\end{figure*}
\begin{figure*}
\includegraphics[width=0.99\textwidth, height=0.18\textheight]{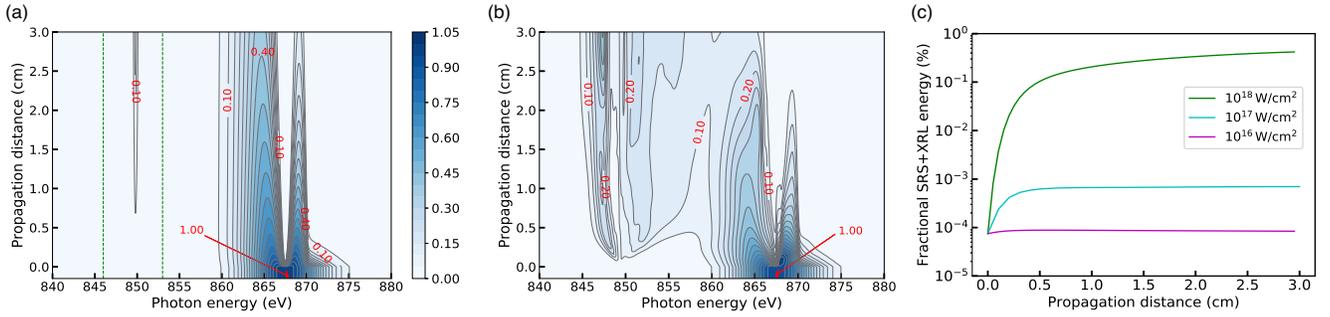}
\caption{\label{fig:SpectralEvolution} Spectral evolution during propagation through the five-level Ne, Ne$^+$ and 1s-continuum system at 1 bar. (a) Spectral evolution of a 0.25 fs, \SI{e18}{W/cm^2} pulse. (b) Spectral evolution of a 0.25 fs, \SI{e19}{W/cm^2} pulse. (c) Growth of the fraction of transmitted energy contained in the SRS+XRL peak (integrated between 846 - 853 eV) for \SI{e16},\SI{e17}, \SI{e18}{W/cm^2} peak intensities.
}
\end{figure*}

\begin{figure*}
\centering
\includegraphics[width=0.99\textwidth, height=0.36\textheight]{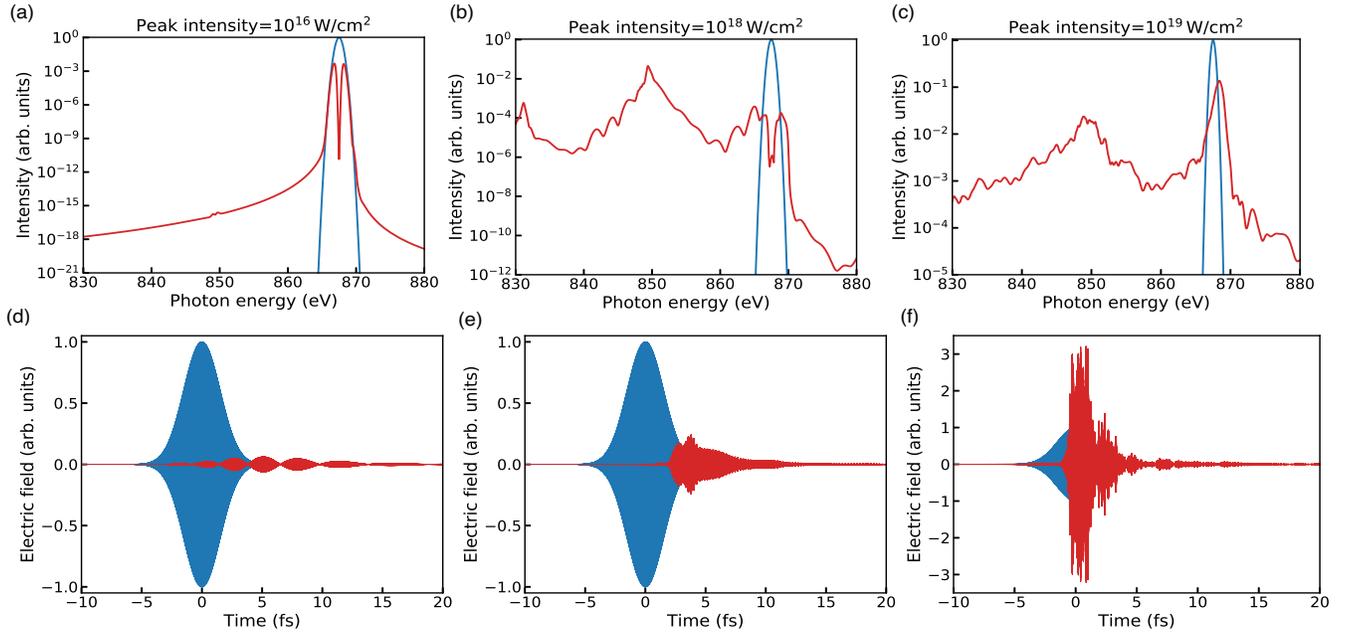}
\caption{\label{fig:intensities_2.5}Spectral and temporal profiles for the incident and transmitted pulse. Incident pulse is 2.5 fs (FWHM) Gaussian centered at 867.5 eV.  Transmission through 3 cm Ne gas at 1 bar pressure calculated using the five-level Ne, Ne$^+$ and 1s-continuum system. Upper row displays spectra for peak intensity (a) \SI{e16}{W/cm^2}, (b) \SI{e18}{W/cm^2} and (c) \SI{e19}{W/cm^2}.  Lower row displays temporal profiles:(d) \SI{e16}{W/cm^2}, (e) \SI{e18}{W/cm^2} and (f) \SI{e19}{W/cm^2} 
%\begin{itemize}
%    \item resonant absorption, the temporal pulse profile is reshaped, oscillating tails appear, the pulse head is compressed
%    \item stimulated Raman occurs at higher intensities, continuum absorption due to photoionization, fine structure is due to the beats between stimulated Raman and main input pulse (one color to two color)
%    \item with even higher intensity, the spectrum changed a lot, the continuum spectrum appears connecting the stimulated Raman and input pulse (explain as dress atom, stimulated AC stark, one color to milt-color)
%    \item the temporal profile also changes, the oscillating tail was modified and the beats between stimulated Raman and input pulse is still there
%\end{itemize}
}
\end{figure*}

Now we consider the temporal and spectral reshaping of x-ray pulses upon propagation through the resonant neon medium as a function of intensity (\SI{e16}-\SI{e19} {W/cm^2}) for two different pulse durations (0.25 fs and 2.5 fs (FWHM).  The \SI{e16} {W/cm^2} represents the linear response of the system and matches simulations at lower intensities (See Fig. \ref{fig:PulseEnergy-3level}c and d for linearity of transmitted pulse energies versus intensity).

Fig. \ref{fig:intensities} displays incident and transmitted spectral and temporal profiles for 0.25-fs Gaussian pulses incident at three peak intensities: \SI{e16}, \SI{e18} and \SI{e19} {W/cm^2}.  At all intensities, there is a sharp dip in the transmitted spectrum at 867.5 eV due to resonant absorption. Photon energies above 870 eV are absorbed via $1s-$photoionization. The ultra-short 0.25 fs pulse corresponds to a transform-limited 7.3 eV (FWHM) bandwidth such that much of the radiation is off resonance and transmitted.  A very small fraction of the incident pulse ($\sim$\SI{e-7}) provides photons at $\sim850$ eV that seed a stimulated Raman transition (SRS) between the core-excited state ($1s^{-1}3p$) and valence-excited state ($2p^{-1}3p$), and on the corresponding x-ray lasing (XRL) transition in the ion ($1s^{-1}$) and ($2p^{-1}$), which is populated via the part of the initial spectrum above 870 eV.
%(We note that the large bandwidth associated with the 0.25 fs pulse overlaps the $1s$ ionization continuum and therefore produces Ne$^+$ that contributes to x-ray lasing, XRL, at a photon energy nearly coincident with that of SRS.) 
Fig.~\ref{fig:SpectralEvolution}(a) and (c) show that at the intermediate intensities \SI{e17}{W/cm^2} and \SI{e18}{W/cm^2}, the stimulated Raman scattering (SRS) signal is exponentially amplified during propagation. At \SI{e18}{W/cm^2}, the spectral strength of the SRS peak is comparable to the peak of the spectrum near 865 eV.  %{\bf LY: Can we be more quantitative about the following comment? We ignored spontaneous emission in our calculations since the field intensity at Raman frequency is much higher than vacuum fluctuation. This should be in the theory section.}
Higher-order scattering processes such as four-wave mixing contribute to a signal at 833 eV, as was observed also in \cite{sun2010propagation}.

\begin{figure*}
\centering
\includegraphics[width=0.99\textwidth, height=0.18\textheight]{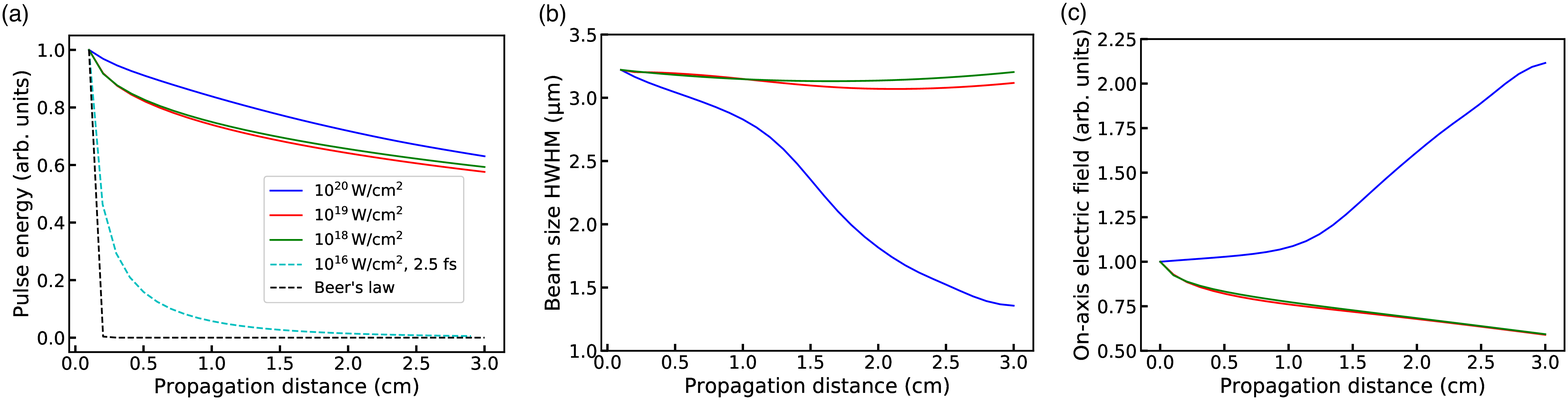}
\caption{\label{fig:sit_selffocus_2level} Pulse propagation through a two-level system (Ne atom $1s$ and $1s^{-1}3p$ without $1s$ photoionization).  (a) Pulse energy attenuation for 0.25-fs pulses with peak intensities between \SI{e18}{W/cm^2} to \SI{e20}{W/cm^2} compared with 2.5-fs pulses at \SI{e18}{W/cm^2} and Beer's law (dashed black line). (b) Beam spot sizes (HWHM) for different peak intensities 0.25-fs pulses as a function of propagation distance. (c) On-axis electric-field strength as a function of propagation distance.
}
\end{figure*}

At even higher intensity, seen in Fig. \ref{fig:intensities}(c), the SRS peak becomes dominant in the transmitted spectrum and spectral content appears throughout the gap between the 867.5 eV resonance and the 849.3 eV SRS peak, Fig. \ref{fig:intensities}c.  We are not aware of this "intermediate" spectrum, between the SRS and resonance peaks,  having been observed before, and we have further investigated its origin. Fig. \ref{fig:SpectralEvolution}(b) shows that the intermediate spectrum appears and then grows exponentially at longer propagation distances ($z>0.3$ cm) as the SRS approaches saturation. While the SRS signal saturates early during propagation also at lower intensities (see Fig. \ref{fig:SpectralEvolution}(c)), it is only the combination of SRS intensity, spectral intensity near the resonance and high fields that gives rise to the intermediate spectrum. At this intensity, the Rabi frequencies associated with both the initial pulse and the generated SRS pulse, and the two resonant transitions, are on the order of 5-10 eV, and the intermediate spectrum can be though of as generalized Rabi sidebands produced by the strong multi-color field. 
%In other words, the presence of two frequencies plus the high fields, which induce Rabi flopping sidebands, gives rise to the "intermediate" spectrum in this combined medium-wave system. 
This interpretation was verified with a separate calculation using two-color, high-intensity (\SI{e19}W/cm$^2$) incident pulses, which was found to give rise to sidebands in the dipole spectrum spanning most of the frequency range between 850 and 867 eV (not shown).  

The temporal pulse profile dramatically differs for the three intensities. At the lowest intensity, the ringing associated with on-resonant pulse propagation for low pulse areas appears.  At the intermediate intensity, finer oscillations associated with the beating between the SRS and main peak are observed. At the highest intensity, \SI{e20} W/cm$^2$, the transmitted pulse is relatively compressed, consistent with the broad spectral profile discussed above, and only a few oscillations of the the low-pulse-area ringing are visible. 

Fig. \ref{fig:intensities_2.5} displays incident and transmitted spectral and temporal profiles for a 2.5-fs Gaussian pulse for the same three intensities.  With the longer pulse, the spectral width (0.73 eV) is comparable to that of the natural width of the absorption resonance (0.24 eV) such that much more of the pulse is absorbed at the low intensity limit -- providing a more efficient energy exchange between the field and medium.  Hence, nonlinearities occur at lower intensity, giving rise to complex spectra. Already by \SI{e18} W/cm$^2$ the SRS resonance peak becomes the largest spectral feature.  At the highest intensity \SI{e19} W/cm$^2$, the rising edge of the pulse is strongly steepened, leading to a much higher peak intensity ($\sim$10-fold), as previously observed computationally for the Ar system \cite{sun2010propagation,sun2009slowdown}.

\subsection{Self-induced transparency and self-focusing\label{sec:results transverse}}

When a weak pulse enters a resonant medium, a fraction of the pulse energy is absorbed by creating excitations. After a few absorption lengths the pulse energy decays to zero according to Beer's law. Self-induced transparency (SIT) refers to the situation when electromagnetic fields pass through a medium with energy attenuation smaller and transit time longer than expected. This happens for a weak pulse with duration shorter than the decoherence timescale \cite{crisp1970propagation,mccall1967self}. The excited dipoles remain in phase collectively after the pulse passes, and thus can radiate power back to the field coherently. As detailed in \cite{crisp1970propagation} the propagation of a small-area pulse in an attenuating medium satisfies an area theorem. That is, the pulse area drops to zero exponentially. However, this does not mean the pulse energy decreases exponentially. The pulse reshapes itself to produce oscillating tails that have alternating phases -- as seen, e.g., in Figs. \ref{fig:population} and \ref{fig:intensities}. The pulse area decreases due to the cancellation between tails, but the pulse energy remains constant. 

\begin{figure*}
\centering
\includegraphics[width=0.99\textwidth, height=0.18\textheight]{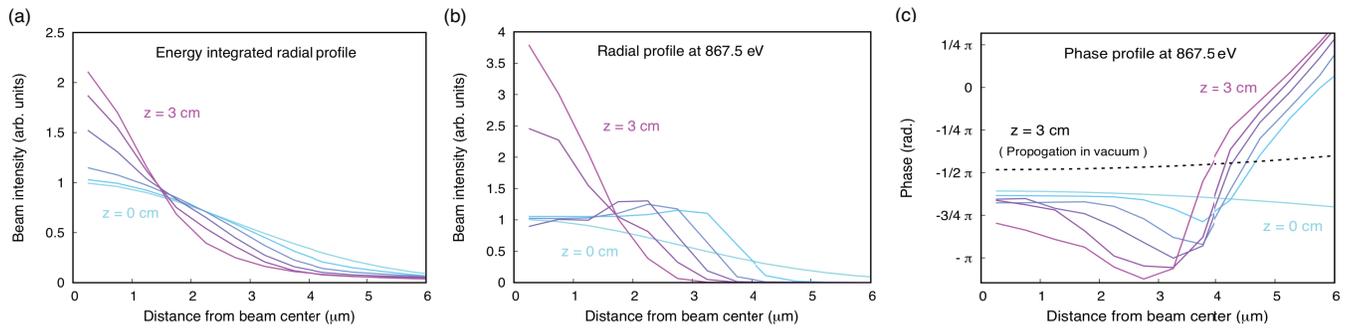}
\caption{\label{fig:wavefront} Pulse propagation through the Ne two-level system for 0.25-fs, \SI{e20}{W/cm^2} pulses: radial intensity and phase profiles for a family of propagation distances from 0 to 3 cm. (a) Radial dependence of the beam intensity integrated over photon energy. (b) Radial dependence of the beam intensity at the resonance energy (867.5 eV) (c) Radial dependence of the phase profile. The black dashed line indicates the output phase profile for a Gaussian beam propagating in vacuum.
%\begin{itemize}
%    \item transverse reshaping
%    \item higher order modes
%    \item wavefront reshaping
%    \item energy flow
%\end{itemize}
}
\end{figure*}

As a preface to our more complex situation we first consider propagation through a simple two-level neon system (1s and 3p) with photoionization ignored as shown in Figs. \ref{fig:sit_selffocus_2level} and \ref{fig:wavefront}.  We first consider the transmitted pulse energy. Fig.\ref{fig:sit_selffocus_2level}(a) shows the evolution of the pulse energy for different incident intensities and two pulse durations (2.5 fs and 0.25 fs). The black dashed line indicates energy absorption according to Beer's law for monochromatic light on resonance. The 2.5 fs pulse at \SI{e18}{W/cm^2} is close to Beer's law, but deviates due to the off-resonant radiation. %The slightly slower decay is due to the ignorance of photoionization effect. 
As is easily understood, the 0.25-fs pulses are attenuated much less than the Beer's law prediction due to substantially more off-resonant components -- short pulses correspond to larger bandwidth and only photons within the absorption bandwidth are absorbed. When the intensity is increased to \SI{e20} W/cm$^2$, population (Rabi) oscillations occur to create the self-induced transparency as shown in Figs. \ref{fig:intensities}a,c. 

Next, with our three-dimensional model we study resonant self-focusing of strong x-ray pulses as shown in Figs. \ref{fig:sit_selffocus_2level}(b),(c). Self-focusing occurs when the electromagnetic field induces a refractive index change in the medium which then reacts back on the field and affects its propagation.  Fig. \ref{fig:sit_selffocus_2level}(b) shows the evolution of beam spot size (HWHM) for different pulse intensities. At \SI{e18}{W/cm^2}, the beam size evolves according to the propagation of a Gaussian beam with 3.2 $\mu$m HWHM at the center of the gas cell. Self-focusing becomes apparent with increasing pulse intensity. The beam radius at \SI{e19}{W/cm^2} deviates slightly from that at \SI{e18}{W/cm^2}; but at \SI{e20}{W/cm^2} there is a dramatic decrease of the beam radius from 3.2 $\mu$m to 1.4 $\mu$m by the end of the gas cell. The corresponding on-axis electric field attenuation is shown in Fig. \ref{fig:sit_selffocus_2level}(c). At low pulse intensity, the on-axis electric field drops, but it grows by a factor of 2.1 at high peak intensity, \SI{e20}{W/cm^2}. 

%{\bf LY:  do we want this paragraph?  If so we probably should expand on the comparison between optical and x-ray Kerr effects.}. Qualitatively, increased absorption at off-axis radial positions where the incident intensity is lower combined with decreased absorption for on-axis light would lead to a decreasing beam spot size as a function of propagation distance. We thus investigated the macroscopic polarization as a function of field amplitude. The medium responds to the incident electromagnetic field through charge displacement which is linearly proportional to the external field at low intensity. When the field is strong enough ($\sim\SI{e19}{W/cm^2}$), the macroscopic polarization becomes a nonlinear function of the applied field. {\bf{Don't get the point here: This agreement verified the self-focusing effect we observed. It is worth stressing that, the self-focusing effect here for a resonant propagation pulse has different origins with the Kerr effect, which is due to the higher-order susceptibility induced by off resonance strong pulse. In fact, due to the rather small x-ray refractive index, the Kerr effect would not be expected to occur at such low x-ray intensities.}}
%////////////\textbf{\textit{Kai:I think we can remove the first part of the paragraph, since it is mainly talking about the same thing as the next paragraph. But we could keep the latter half of this paragraph, which stress that here the self-focusing is on resonance, instead of the normal off-resonant optical Keer effects. I moved it to the end of the next two paragraph.}}

To illustrate the mechanism behind the resonant x-ray self-focusing we show the evolution of the beam's radial and phase profile in Fig. \ref{fig:wavefront} for \SI{e20}{W/cm^2} where a large self-focusing effect is observed. The energy-integrated transverse beam profile is shown in Fig. \ref{fig:wavefront}a for a family of propagation distances, z=0-3 cm. The transverse beam profile is reshaped to a sharp peak in the center with the maximum field strength increased by a factor of 2$\times$ by the end of the gas cell. 

The radial amplitude and phase profiles at the resonance energy, 867.5 eV, are shown in Figs. \ref{fig:wavefront}(b) and (c), respectively. The Gaussian beam profile develops higher-order transverse modes during propagation and the evolution is not as smooth as that of the energy-integrated profile. Rather than monotonically decreasing, the beam first flattens and grows, and then shrinks to a very sharp peak with a 4-fold increased maximum field by the end of the cell. 

The evolution of the phase profile of the beam is shown in Fig.\ref{fig:wavefront}(c). For reference, the phase profile of a Gaussian beam, focused at z=1.5 cm, after propagation in vacuum for z=3 cm is shown as a black dashed line.  This beam is defocusing at z=3, with an increasing phase as a function of radial position.  For the beam propagating through Ne gas, initially the beam is focusing toward the center of the cell and shows a slightly decreasing phase for larger radial distances.  On-axis, the relatively high-intensity creates stronger SIT, while the off-axis fields are attenuated. There is a reshaping of the wavefront to be strongly focusing which induces energy flow inward and growth of the on-axis power density. Modification of absorption and phase due to changes of the imaginary and real parts of the nonlinear complex refractive index contribute to this pulse front reshaping. In summary, resonant x-ray self-focusing results from the lower intensity off-axis fields interacting differently than the high intensity on-axis fields as observed in the optical regime \cite{wright1973self,gibbs1976coherent,marburger1975self}.  It is worth noting that, the resonant self-focusing effect here has a different origin compared to the normal off-resonant optical Kerr effect, which results from higher-order susceptibility induced by strong fields. In fact, due to the rather small x-ray refractive index, the off-resonant Kerr effect would not be expected to occur until reaching much higher x-ray intensities.

\begin{figure*} [!htbp!]
\centering
\includegraphics[width=0.74\textwidth, height=0.42\textheight]{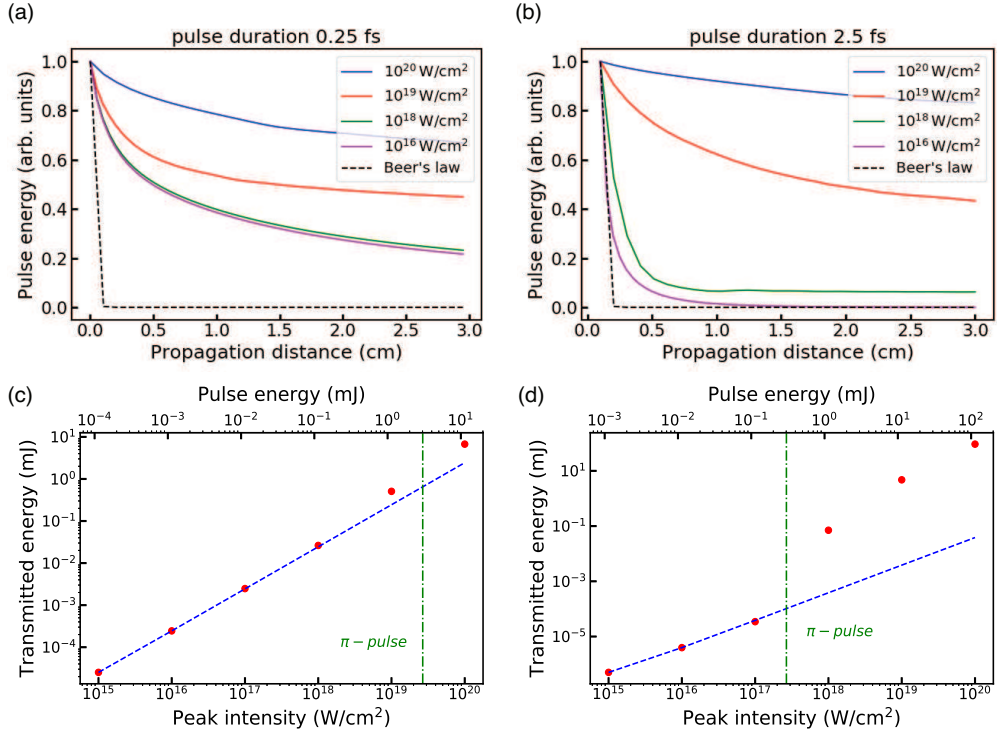}
\caption{\label{fig:PulseEnergy-3level} Transmitted pulse energies for 0.25 and 2.5-fs pulses five-level Ne system with 1s photoionization. The dashed blue lines in (c) and (d) represent a linear relationship between the incident and transmitted pulse energies, i.e slope = 1.0. (a) Propagation of 0.25-fs pulse with peak intensities from \SI{e16}{W/cm^2} to \SI{e20}{W/cm^2} (b) Propagation of 2.5-fs pulse with peak intensities \SI{e16}{W/cm^2} to \SI{e20}{W/cm^2}. (c) Transmitted pulse energy after passage through 3 cm gas at various intensities for 0.25 fs pulses. The blue dashed line has a slope of one. (d) Same as (c) for 2.5 fs pulses.
%\begin{itemize}
%    \item diffraction appears at low pulse intensities, the beam size (FWHM) begin decreases when pulse energy is increases ($\pi/2$ pulse), self-focusing occurs when the intensity is even higher ($2\pi$ pulse)
%    \item Self-focusing effect is different between 2-levels and 3-levels system. The effect happens at lower pulse intensities and it is stronger in 3-levels system. 
%    \item Instead of focusing to a singular point, it saturates at some points and begins to defocus.  
%    \item note that the profile is not necessary to be gaussian anymore, multiple transverse modes 
%    \item the on axis energy density intensity increase as the self-focusing happens
%\end{itemize}
}
\end{figure*}

\begin{figure*} [!htbp!]
\centering
\includegraphics[width=0.74\textwidth, height=0.21\textheight]{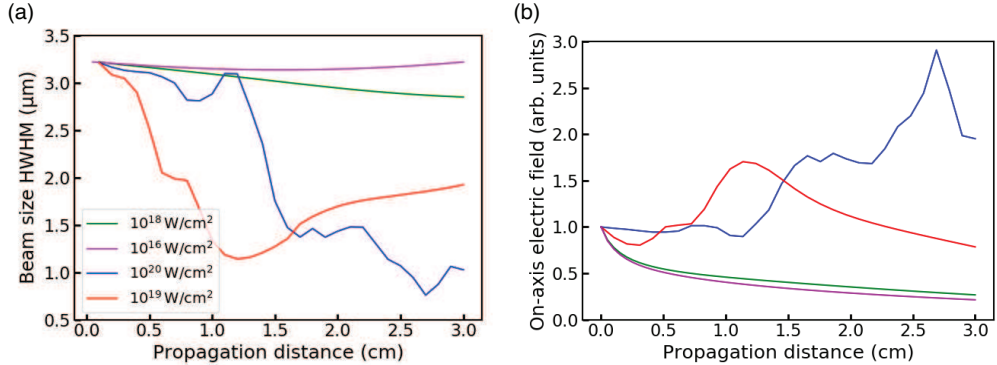}
\caption{\label{fig:wavefront-3level} (a) Beam spot size of x-ray pulses with 0.25 fs pulse duration and different peak intensities ranging from \SI{e16}{W/cm^2} to \SI{e20}{W/cm^2} propagating through the five-level Ne, Ne$^+$ and 1s-continuum system. (b) The evolution of the corresponding on-axis power densities is shown in (b).
%\begin{itemize}
%    \item diffraction appears at low pulse intensities, the beam size (FWHM) begin decreases when pulse energy is increases ($\pi/2$ pulse), self-focusing occurs when the intensity is even higher ($2\pi$ pulse)
%    \item Self-focusing effect is different between 2-levels and 3-levels system. The effect happens at lower pulse intensities and it is stronger in 3-levels system. 
%    \item Instead of focusing to a singular point, it saturates at some points and begins to defocus.  
%    \item note that the profile is not necessary to be gaussian anymore, multiple transverse modes 
%    \item the on axis energy density intensity increase as the self-focusing happens
%\end{itemize}
}
\end{figure*}

 For the five-level Ne, Ne$^+$ and 1s-continuum system the SIT and self-focusing display a much more complicated dependence on intensity and pulse duration. The attenuation of an ultra-short 0.25 fs pulse for several intensities are shown in Fig. \ref{fig:PulseEnergy-3level}(a). For peak intensities lower than \SI{e18}{W/cm^2} stimulated Raman scattering is weak; the five-level system behaves similar to a two-level system, where non-Beer's law behavior is due to the ultra-short pulse duration. One difference is that $1s$-photoionization causes the transmission to drop from 0.6 to 0.2. For the 2.5-fs pulse propagating through the five-level system, (Fig. \ref{fig:PulseEnergy-3level}(b), deviations occur at relatively low intensities.  Comparison of the transmitted pulse energies as a function of incident intensity for 0.25- and 2.5-fs pulses are shown in Figs. \ref{fig:PulseEnergy-3level}(c) and (d).  A dashed line is shown for the intensity associated with a $\pi$-pulse for the $1s\rightarrow3p$ transition.  As can be seen from the Fig. \ref{fig:PulseEnergy-3level}, deviations from linear transmission occur at intensities greater than that required for a $\pi$-pulse -- at the intensities associated with SIT and self-focusing. % {\bf Spectral analysis indicates that a substantial fraction of the pulse energy is transformed into stimulated Raman signal in the three-level system, thus suppressing resonant absorption.}
%{\bf Compared to the two-level system, SIT is more sensitive to pulse intensity in the three-level system. More energy penetrates the gas as higher pulse intensities induce stronger SRS signal. Moreover, at high intensities, more radiation energy is stored in the intermediate frequencies between the Raman and resonant signal which further suppresses resonant absorption.}

We now turn our attention to self-focusing in the five-level system. At low intensities, the beam spot size for a five-level system, Fig. \ref{fig:wavefront-3level}(a), behaves the same as in a two-level system, Fig. \ref{fig:wavefront}(b). However at \SI{e19}{W/cm^2} beam size oscillations appear and a stronger self-focusing effect occurs. Surprisingly, the beam radius due to self-focusing does not necessarily decrease more rapidly with higher peak intensity - compare \SI{e19}{W/cm^2} and \SI{e20}{W/cm^2}. The x-ray radial evolution is clearly coupled to changes in the on-axis power density shown in Fig. \ref{fig:wavefront-3level}(b). This non-monotonic behavior is the sign of highly nonlinear radiation-matter interactions. The strong interaction between the field and atom ensemble gives rise to more complex electronic transitions within the dressed atomic states which further induces polarization fluctuations and modifies the refractive index. Higher field strengths in the center of the beam lead to stronger SRS and less absorption.  This mechanism exacerbates the self-focusing effect at \SI{e20}{W/cm^2} leading to beam spot shrinkage to one third of the original size and an on-axis field strength 2.8$\times$ that of the incident pulse. 

\subsection{SASE pulse propagation}

\begin{figure*} [!htbp!]
\centering
\includegraphics[width=0.74\textwidth, height=0.21\textheight]{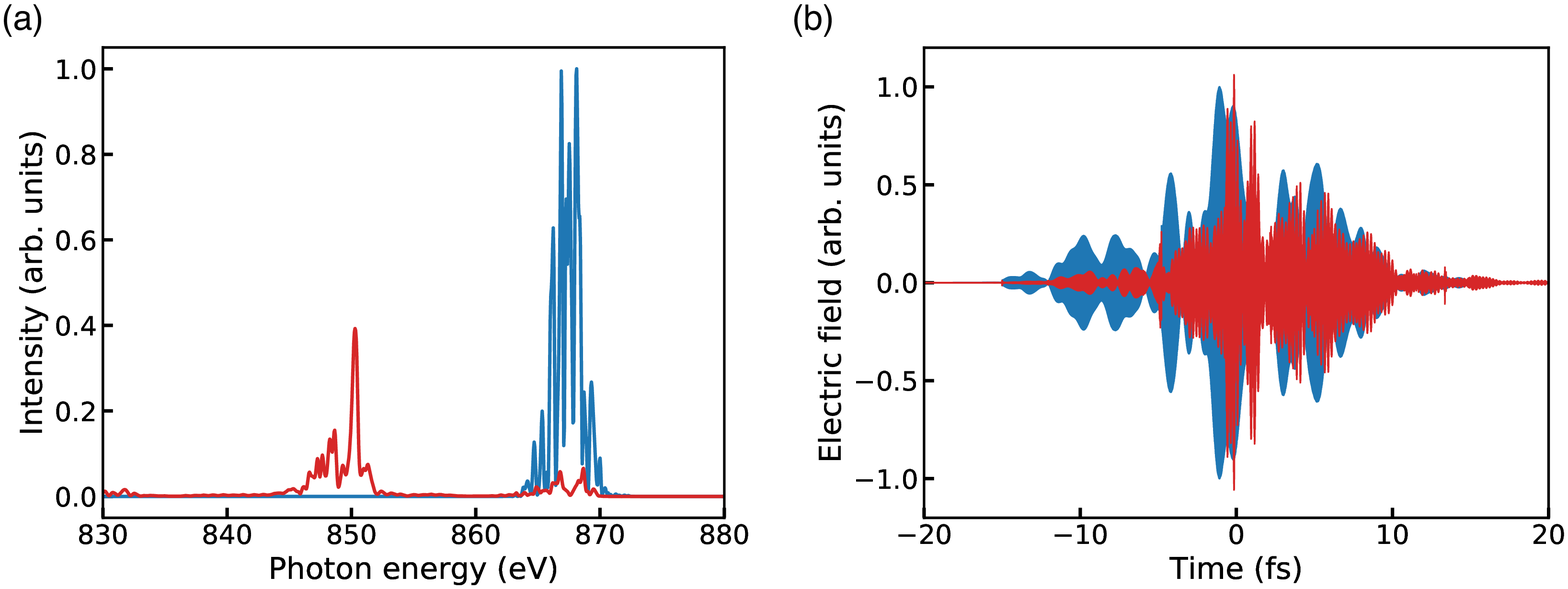}
\caption{\label{fig:SASE} (a) Input (blue) and output (red) spectrum of SASE pulses with 10 fs (FWHM) pulse duration and peak intensity \SI{e19}{W/cm^2} propagating through the five-level system with photoionization. The beam spot size is 2 $\mu$m (FWHM) and pulse energy is 5 mJ. (b) The corresponding input and output x-ray temporal profile.
}
\end{figure*}

It is natural to compare the propagation of a Gaussian with a SASE pulse, since those are most readily available at XFELs. We simulate SASE pulses with typical parameters of $\tau=10$ fs pulse duration, bandwidth = $\Delta E/E=0.005$, and peak intensity of \SI{e19}{W/cm^2}. These parameters correspond to a beam spot size of 2 $\mu$m, and pulse energy of 5 mJ.  The pulse is produced by a computer simulation technique developed for generating superposed coherent and chaotic radiation \cite{vannucci1980computer}. The radiation field amplitudes at different frequency can be modeled as chaotic variables with Gaussian statistics. The ensemble averaged SASE pulse is assumed to have a Gaussian power spectrum with bandwidth $\frac{\Delta E}{E}$ of the central photon energy. The field amplitude at different frequency are independent zero-mean Gaussian random variables whose variance is proportional to the corresponding power spectrum. The SASE pulse is first generated in the energy domain with 4096 points spaced by 0.04 eV. %The energy step is 0.04 eV, which corresponding to a 100 fs time window in time domain.  
After the inverse Fourier transform to the time domain and normalization, a Gaussian envelope is used to create the pulse. This field amplitude and phase are then linearly interpolated to 524288 grid points to get the input field for the TDSE-MWE propagation code. As shown in the Fig. \ref{fig:SASE}, the generated SASE pulse is comprised of coherent spikes with $\tau_{coh}=\frac{h}{\Delta E}\sim1$ fs duration, which is related to the total spectral bandwidth. The averaged 0.4 eV width spikes in the SASE energy spectrum correspond to $\frac{h}{\tau}$.

The simulation results for propagation of the SASE pulse shown in Fig. \ref{fig:SASE} have features that closely resemble the Gaussian pulse. The input SASE x-ray pulse is absorbed and the SRS signal is generated at 850 eV. The SRS saturates upon propagation to 0.5 cm, and the SRS spectrum contains spikes associated with the incident SASE spectrum. Four-wave mixing signals with $\sim10\times$ smaller intensity are observed at 830 eV. The SASE pulse is delayed and reshaped (compare the red and blue pulse envelopes in Fig. \ref{fig:SASE}b). Some very short $\sim20$ as (FWHM) bursts are generated on top of the SASE pulse.

The atomic inner-shell x-ray laser \cite{rohringer2012atomic} was also simulated using our code by tuning the incident energy above the ionization threshold. The simulation was carried out for a SASE pulse with 40-fs duration and 960-eV central photon energy propagating through a 500 torr Neon gas. The 0.3 mJ x-ray pulse was focused to 2 $\mu$m spot size to produce peak intensity around \SI{e17}{W/cm^2}. These parameters correspond to the experiment performed at LCLS \cite{rohringer2012atomic,weninger2014transient}. We calculate a saturation length of 6 mm and 0.4\% fractional XRL energy in agreement with the experiment. 

\section{summary\label{sec:conclusion}}
In this paper, we investigated resonant propagation of ultra-short, high-intensity XFEL pulses in a gaseous medium from the linear to the nonlinear regimes. We solved the three-dimensional time-dependent Schr\"odinger equation (TDSE) for the single atom response and the Maxwell wave equation (MWE) . Specifically, the propagation of XFEL pulses with photon energy resonant with the Ne $1s\rightarrow3p$ transition through an optically thick target was investigated.  The stimulated Raman scattering (SRS) signal, i.e. the transition between core-excited ($1s^{-1}3p$) and valence-excited state ($2p^{-1}3p$), grows exponentially during propagation for intensities up to \SI{e18}{W/cm^2} for 0.25-fs pulses. At higher intensities, spectral intensity not associated with atomic transitions appear during propagation due to the existence of two ingredients: significant SRS intensity and strong fields to induce Rabi-flopping sidebands.  Stronger interactions occur with the longer 2.5-fs pulse, leading to strong spectral and temporal reshaping of a XFEL pulse.  X-ray self-induced transparency and self-focusing are observed when the intensity is sufficient to induce a $\pi$-pulse on the resonant transition.  The newly developed TDSE-MWE methodology is very general: readily applicable to SASE pulses and demonstrated to reproduce the off-resonant propagation that leads to the atomic x-ray laser \cite{weninger2013PRA,rohringer2012atomic}. Generalizing our formalism to an arbitrarily polarized field would be simple at the level of the MWE, where each polarization component would propagate independently.  Generalizing the single-atom response would involve including more states, with different m-quantum numbers, and incorporating selection rules. Such extensions would be of interest for ultrafast circular dichroism studies. In summary, the understanding of resonant propagation at high x-ray intensities has potential applications for XFEL pulse shaping and is relevant for x-ray optics and transient absorption spectroscopy. An experiment devoted to quantifying these effects is feasible at various XFEL facilities by measuring the energy spectrum before and after propagation through the target gas. We anticipate investigating nonlinear resonant propagation effects experimentally in the near future.

%\begin{itemize}
%    \item 3D x-ray resonant propagation simulation in real gas cell (1 bar, cm) including Auger decay, photoionization.
%    \item experiments can be done
%    \item new phenomina occurs at high intensity x-ray short pulses. the efficiently populates core-excited states, Rabi oscillation before decoherent decay, and Raman scattering  
%    \item We observed stimulated Raman scattering, SIT, Self-focusing and Stimulated AC stark. (nonlinear phenomena in x-ray regime, longitudinal and transverse reshaping) 
%\end{itemize}

\begin{acknowledgements}
L.Y. and K.L. are grateful for stimulating initial discussions with N. Rohringer and V. Majety.  This work was supported by the U.S. Department of Energy, Office of Science, Basic Energy Science, Chemical Sciences, Geosciences and Biosciences Division that supported the Argonne group under contract number DE-AC02-06CH11357, and the LSU group under contract number DE-SC0010431.
\end{acknowledgements}
\bibliography{ResPropXray}

\begin{thebibliography}{45}%
\makeatletter
\providecommand \@ifxundefined [1]{%
 \@ifx{#1\undefined}
}%
\providecommand \@ifnum [1]{%
 \ifnum #1\expandafter \@firstoftwo
 \else \expandafter \@secondoftwo
 \fi
}%
\providecommand \@ifx [1]{%
 \ifx #1\expandafter \@firstoftwo
 \else \expandafter \@secondoftwo
 \fi
}%
\providecommand \natexlab [1]{#1}%
\providecommand \enquote  [1]{``#1''}%
\providecommand \bibnamefont  [1]{#1}%
\providecommand \bibfnamefont [1]{#1}%
\providecommand \citenamefont [1]{#1}%
\providecommand \href@noop [0]{\@secondoftwo}%
\providecommand \href [0]{\begingroup \@sanitize@url \@href}%
\providecommand \@href[1]{\@@startlink{#1}\@@href}%
\providecommand \@@href[1]{\endgroup#1\@@endlink}%
\providecommand \@sanitize@url [0]{\catcode `\\12\catcode `\$12\catcode
  `\&12\catcode `\#12\catcode `\^12\catcode `\_12\catcode `\%12\relax}%
\providecommand \@@startlink[1]{}%
\providecommand \@@endlink[0]{}%
\providecommand \url  [0]{\begingroup\@sanitize@url \@url }%
\providecommand \@url [1]{\endgroup\@href {#1}{\urlprefix }}%
\providecommand \urlprefix  [0]{URL }%
\providecommand \Eprint [0]{\href }%
\providecommand \doibase [0]{https://doi.org/}%
\providecommand \selectlanguage [0]{\@gobble}%
\providecommand \bibinfo  [0]{\@secondoftwo}%
\providecommand \bibfield  [0]{\@secondoftwo}%
\providecommand \translation [1]{[#1]}%
\providecommand \BibitemOpen [0]{}%
\providecommand \bibitemStop [0]{}%
\providecommand \bibitemNoStop [0]{.\EOS\space}%
\providecommand \EOS [0]{\spacefactor3000\relax}%
\providecommand \BibitemShut  [1]{\csname bibitem#1\endcsname}%
\let\auto@bib@innerbib\@empty
%</preamble>
\bibitem [{\citenamefont {Emma}\ \emph {et~al.}(2010)\citenamefont {Emma},
  \citenamefont {Akre}, \citenamefont {Arthur}, \citenamefont {Bionta},
  \citenamefont {Bostedt}, \citenamefont {Bozek}, \citenamefont {Brachmann},
  \citenamefont {Bucksbaum}, \citenamefont {Coffee}, \citenamefont {Decker}
  \emph {et~al.}}]{emma2010first}%
  \BibitemOpen
  \bibfield  {author} {\bibinfo {author} {\bibfnamefont {P.}~\bibnamefont
  {Emma}}, \bibinfo {author} {\bibfnamefont {R.}~\bibnamefont {Akre}}, \bibinfo
  {author} {\bibfnamefont {J.}~\bibnamefont {Arthur}}, \bibinfo {author}
  {\bibfnamefont {R.}~\bibnamefont {Bionta}}, \bibinfo {author} {\bibfnamefont
  {C.}~\bibnamefont {Bostedt}}, \bibinfo {author} {\bibfnamefont
  {J.}~\bibnamefont {Bozek}}, \bibinfo {author} {\bibfnamefont
  {A.}~\bibnamefont {Brachmann}}, \bibinfo {author} {\bibfnamefont
  {P.}~\bibnamefont {Bucksbaum}}, \bibinfo {author} {\bibfnamefont
  {R.}~\bibnamefont {Coffee}}, \bibinfo {author} {\bibfnamefont {F.-J.}\
  \bibnamefont {Decker}}, \emph {et~al.},\ }\bibfield  {title} {\bibinfo
  {title} {First lasing and operation of an {\aa}ngstrom-wavelength
  free-electron laser},\ }\href@noop {} {\bibfield  {journal} {\bibinfo
  {journal} {nature photonics}\ }\textbf {\bibinfo {volume} {4}},\ \bibinfo
  {pages} {641} (\bibinfo {year} {2010})}\BibitemShut {NoStop}%
\bibitem [{\citenamefont {Ishikawa}\ \emph {et~al.}(2012)\citenamefont
  {Ishikawa}, \citenamefont {Aoyagi}, \citenamefont {Asaka}, \citenamefont
  {Asano}, \citenamefont {Azumi}, \citenamefont {Bizen}, \citenamefont {Ego},
  \citenamefont {Fukami}, \citenamefont {Fukui}, \citenamefont {Furukawa} \emph
  {et~al.}}]{ishikawa2012compact}%
  \BibitemOpen
  \bibfield  {author} {\bibinfo {author} {\bibfnamefont {T.}~\bibnamefont
  {Ishikawa}}, \bibinfo {author} {\bibfnamefont {H.}~\bibnamefont {Aoyagi}},
  \bibinfo {author} {\bibfnamefont {T.}~\bibnamefont {Asaka}}, \bibinfo
  {author} {\bibfnamefont {Y.}~\bibnamefont {Asano}}, \bibinfo {author}
  {\bibfnamefont {N.}~\bibnamefont {Azumi}}, \bibinfo {author} {\bibfnamefont
  {T.}~\bibnamefont {Bizen}}, \bibinfo {author} {\bibfnamefont
  {H.}~\bibnamefont {Ego}}, \bibinfo {author} {\bibfnamefont {K.}~\bibnamefont
  {Fukami}}, \bibinfo {author} {\bibfnamefont {T.}~\bibnamefont {Fukui}},
  \bibinfo {author} {\bibfnamefont {Y.}~\bibnamefont {Furukawa}}, \emph
  {et~al.},\ }\bibfield  {title} {\bibinfo {title} {A compact x-ray
  free-electron laser emitting in the sub-{\aa}ngstr{\"o}m region},\
  }\href@noop {} {\bibfield  {journal} {\bibinfo  {journal} {nature photonics}\
  }\textbf {\bibinfo {volume} {6}},\ \bibinfo {pages} {540} (\bibinfo {year}
  {2012})}\BibitemShut {NoStop}%
\bibitem [{\citenamefont {Young}\ \emph {et~al.}(2010)\citenamefont {Young},
  \citenamefont {Kanter}, \citenamefont {Kraessig}, \citenamefont {Li},
  \citenamefont {March}, \citenamefont {Pratt}, \citenamefont {Santra},
  \citenamefont {Southworth}, \citenamefont {Rohringer}, \citenamefont
  {DiMauro} \emph {et~al.}}]{young2010femtosecond}%
  \BibitemOpen
  \bibfield  {author} {\bibinfo {author} {\bibfnamefont {L.}~\bibnamefont
  {Young}}, \bibinfo {author} {\bibfnamefont {E.}~\bibnamefont {Kanter}},
  \bibinfo {author} {\bibfnamefont {B.}~\bibnamefont {Kraessig}}, \bibinfo
  {author} {\bibfnamefont {Y.}~\bibnamefont {Li}}, \bibinfo {author}
  {\bibfnamefont {A.}~\bibnamefont {March}}, \bibinfo {author} {\bibfnamefont
  {S.}~\bibnamefont {Pratt}}, \bibinfo {author} {\bibfnamefont
  {R.}~\bibnamefont {Santra}}, \bibinfo {author} {\bibfnamefont
  {S.}~\bibnamefont {Southworth}}, \bibinfo {author} {\bibfnamefont
  {N.}~\bibnamefont {Rohringer}}, \bibinfo {author} {\bibfnamefont
  {L.}~\bibnamefont {DiMauro}}, \emph {et~al.},\ }\bibfield  {title} {\bibinfo
  {title} {Femtosecond electronic response of atoms to ultra-intense x-rays},\
  }\href@noop {} {\bibfield  {journal} {\bibinfo  {journal} {Nature}\ }\textbf
  {\bibinfo {volume} {466}},\ \bibinfo {pages} {56} (\bibinfo {year}
  {2010})}\BibitemShut {NoStop}%
\bibitem [{\citenamefont {Ho}\ \emph {et~al.}(2014)\citenamefont {Ho},
  \citenamefont {Bostedt}, \citenamefont {Schorb},\ and\ \citenamefont
  {Young}}]{ho2014theoretical}%
  \BibitemOpen
  \bibfield  {author} {\bibinfo {author} {\bibfnamefont {P.~J.}\ \bibnamefont
  {Ho}}, \bibinfo {author} {\bibfnamefont {C.}~\bibnamefont {Bostedt}},
  \bibinfo {author} {\bibfnamefont {S.}~\bibnamefont {Schorb}},\ and\ \bibinfo
  {author} {\bibfnamefont {L.}~\bibnamefont {Young}},\ }\bibfield  {title}
  {\bibinfo {title} {Theoretical tracking of resonance-enhanced multiple
  ionization pathways in x-ray free-electron laser pulses},\ }\href@noop {}
  {\bibfield  {journal} {\bibinfo  {journal} {Physical review letters}\
  }\textbf {\bibinfo {volume} {113}},\ \bibinfo {pages} {253001} (\bibinfo
  {year} {2014})}\BibitemShut {NoStop}%
\bibitem [{\citenamefont {Rudek}\ \emph {et~al.}(2012)\citenamefont {Rudek},
  \citenamefont {Son}, \citenamefont {Foucar}, \citenamefont {Epp},
  \citenamefont {Erk}, \citenamefont {Hartmann}, \citenamefont {Adolph},
  \citenamefont {Andritschke}, \citenamefont {Aquila}, \citenamefont {Berrah},
  \citenamefont {Bostedt}, \citenamefont {Bozek}, \citenamefont {Coppola},
  \citenamefont {Filsinger}, \citenamefont {Gorke}, \citenamefont {Gorkhover},
  \citenamefont {Graafsma}, \citenamefont {Gumprecht}, \citenamefont
  {Hartmann}, \citenamefont {Hauser}, \citenamefont {Herrmann}, \citenamefont
  {Hirsemann}, \citenamefont {Holl}, \citenamefont {H{\"o}mke}, \citenamefont
  {Journel}, \citenamefont {Kaiser}, \citenamefont {Kimmel}, \citenamefont
  {Krasniqi}, \citenamefont {K{\"u}hnel}, \citenamefont {Matysek},
  \citenamefont {Messerschmidt}, \citenamefont {Miesner}, \citenamefont
  {M{\"o}ller}, \citenamefont {Moshammer}, \citenamefont {Nagaya},
  \citenamefont {Nilsson}, \citenamefont {Potdevin}, \citenamefont
  {Pietschner}, \citenamefont {Reich}, \citenamefont {Rupp}, \citenamefont
  {Schaller}, \citenamefont {Schlichting}, \citenamefont {Schmidt},
  \citenamefont {Schopper}, \citenamefont {Schorb}, \citenamefont
  {Schr{\"o}ter}, \citenamefont {Schulz}, \citenamefont {Simon}, \citenamefont
  {Soltau}, \citenamefont {Str{\"u}der}, \citenamefont {Ueda}, \citenamefont
  {Weidenspointner}, \citenamefont {Santra}, \citenamefont {Ullrich},
  \citenamefont {Rudenko},\ and\ \citenamefont {Rolles}}]{Rudek2012NatPho}%
  \BibitemOpen
  \bibfield  {author} {\bibinfo {author} {\bibfnamefont {B.}~\bibnamefont
  {Rudek}}, \bibinfo {author} {\bibfnamefont {S.-K.}\ \bibnamefont {Son}},
  \bibinfo {author} {\bibfnamefont {L.}~\bibnamefont {Foucar}}, \bibinfo
  {author} {\bibfnamefont {S.~W.}\ \bibnamefont {Epp}}, \bibinfo {author}
  {\bibfnamefont {B.}~\bibnamefont {Erk}}, \bibinfo {author} {\bibfnamefont
  {R.}~\bibnamefont {Hartmann}}, \bibinfo {author} {\bibfnamefont
  {M.}~\bibnamefont {Adolph}}, \bibinfo {author} {\bibfnamefont
  {R.}~\bibnamefont {Andritschke}}, \bibinfo {author} {\bibfnamefont
  {A.}~\bibnamefont {Aquila}}, \bibinfo {author} {\bibfnamefont
  {N.}~\bibnamefont {Berrah}}, \bibinfo {author} {\bibfnamefont
  {C.}~\bibnamefont {Bostedt}}, \bibinfo {author} {\bibfnamefont
  {J.}~\bibnamefont {Bozek}}, \bibinfo {author} {\bibfnamefont
  {N.}~\bibnamefont {Coppola}}, \bibinfo {author} {\bibfnamefont
  {F.}~\bibnamefont {Filsinger}}, \bibinfo {author} {\bibfnamefont
  {H.}~\bibnamefont {Gorke}}, \bibinfo {author} {\bibfnamefont
  {T.}~\bibnamefont {Gorkhover}}, \bibinfo {author} {\bibfnamefont
  {H.}~\bibnamefont {Graafsma}}, \bibinfo {author} {\bibfnamefont
  {L.}~\bibnamefont {Gumprecht}}, \bibinfo {author} {\bibfnamefont
  {A.}~\bibnamefont {Hartmann}}, \bibinfo {author} {\bibfnamefont
  {G.}~\bibnamefont {Hauser}}, \bibinfo {author} {\bibfnamefont
  {S.}~\bibnamefont {Herrmann}}, \bibinfo {author} {\bibfnamefont
  {H.}~\bibnamefont {Hirsemann}}, \bibinfo {author} {\bibfnamefont
  {P.}~\bibnamefont {Holl}}, \bibinfo {author} {\bibfnamefont {A.}~\bibnamefont
  {H{\"o}mke}}, \bibinfo {author} {\bibfnamefont {L.}~\bibnamefont {Journel}},
  \bibinfo {author} {\bibfnamefont {C.}~\bibnamefont {Kaiser}}, \bibinfo
  {author} {\bibfnamefont {N.}~\bibnamefont {Kimmel}}, \bibinfo {author}
  {\bibfnamefont {F.}~\bibnamefont {Krasniqi}}, \bibinfo {author}
  {\bibfnamefont {K.-U.}\ \bibnamefont {K{\"u}hnel}}, \bibinfo {author}
  {\bibfnamefont {M.}~\bibnamefont {Matysek}}, \bibinfo {author} {\bibfnamefont
  {M.}~\bibnamefont {Messerschmidt}}, \bibinfo {author} {\bibfnamefont
  {D.}~\bibnamefont {Miesner}}, \bibinfo {author} {\bibfnamefont
  {T.}~\bibnamefont {M{\"o}ller}}, \bibinfo {author} {\bibfnamefont
  {R.}~\bibnamefont {Moshammer}}, \bibinfo {author} {\bibfnamefont
  {K.}~\bibnamefont {Nagaya}}, \bibinfo {author} {\bibfnamefont
  {B.}~\bibnamefont {Nilsson}}, \bibinfo {author} {\bibfnamefont
  {G.}~\bibnamefont {Potdevin}}, \bibinfo {author} {\bibfnamefont
  {D.}~\bibnamefont {Pietschner}}, \bibinfo {author} {\bibfnamefont
  {C.}~\bibnamefont {Reich}}, \bibinfo {author} {\bibfnamefont
  {D.}~\bibnamefont {Rupp}}, \bibinfo {author} {\bibfnamefont {G.}~\bibnamefont
  {Schaller}}, \bibinfo {author} {\bibfnamefont {I.}~\bibnamefont
  {Schlichting}}, \bibinfo {author} {\bibfnamefont {C.}~\bibnamefont
  {Schmidt}}, \bibinfo {author} {\bibfnamefont {F.}~\bibnamefont {Schopper}},
  \bibinfo {author} {\bibfnamefont {S.}~\bibnamefont {Schorb}}, \bibinfo
  {author} {\bibfnamefont {C.-D.}\ \bibnamefont {Schr{\"o}ter}}, \bibinfo
  {author} {\bibfnamefont {J.}~\bibnamefont {Schulz}}, \bibinfo {author}
  {\bibfnamefont {M.}~\bibnamefont {Simon}}, \bibinfo {author} {\bibfnamefont
  {H.}~\bibnamefont {Soltau}}, \bibinfo {author} {\bibfnamefont
  {L.}~\bibnamefont {Str{\"u}der}}, \bibinfo {author} {\bibfnamefont
  {K.}~\bibnamefont {Ueda}}, \bibinfo {author} {\bibfnamefont {G.}~\bibnamefont
  {Weidenspointner}}, \bibinfo {author} {\bibfnamefont {R.}~\bibnamefont
  {Santra}}, \bibinfo {author} {\bibfnamefont {J.}~\bibnamefont {Ullrich}},
  \bibinfo {author} {\bibfnamefont {A.}~\bibnamefont {Rudenko}},\ and\ \bibinfo
  {author} {\bibfnamefont {D.}~\bibnamefont {Rolles}},\ }\bibfield  {title}
  {\bibinfo {title} {Ultra-efficient ionization of heavy atoms by intense x-ray
  free-electron laser pulses},\ }\href
  {https://doi.org/10.1038/nphoton.2012.261} {\bibfield  {journal} {\bibinfo
  {journal} {Nature Photonics}\ }\textbf {\bibinfo {volume} {6}},\ \bibinfo
  {pages} {858} (\bibinfo {year} {2012})}\BibitemShut {NoStop}%
\bibitem [{\citenamefont {Rudenko}\ \emph {et~al.}(2017)\citenamefont
  {Rudenko}, \citenamefont {Inhester}, \citenamefont {Hanasaki}, \citenamefont
  {Li}, \citenamefont {Robatjazi}, \citenamefont {Erk}, \citenamefont {Boll},
  \citenamefont {Toyota}, \citenamefont {Hao}, \citenamefont {Vendrell} \emph
  {et~al.}}]{rudenko2017femtosecond}%
  \BibitemOpen
  \bibfield  {author} {\bibinfo {author} {\bibfnamefont {A.}~\bibnamefont
  {Rudenko}}, \bibinfo {author} {\bibfnamefont {L.}~\bibnamefont {Inhester}},
  \bibinfo {author} {\bibfnamefont {K.}~\bibnamefont {Hanasaki}}, \bibinfo
  {author} {\bibfnamefont {X.}~\bibnamefont {Li}}, \bibinfo {author}
  {\bibfnamefont {S.}~\bibnamefont {Robatjazi}}, \bibinfo {author}
  {\bibfnamefont {B.}~\bibnamefont {Erk}}, \bibinfo {author} {\bibfnamefont
  {R.}~\bibnamefont {Boll}}, \bibinfo {author} {\bibfnamefont {K.}~\bibnamefont
  {Toyota}}, \bibinfo {author} {\bibfnamefont {Y.}~\bibnamefont {Hao}},
  \bibinfo {author} {\bibfnamefont {O.}~\bibnamefont {Vendrell}}, \emph
  {et~al.},\ }\bibfield  {title} {\bibinfo {title} {Femtosecond response of
  polyatomic molecules to ultra-intense hard x-rays},\ }\href@noop {}
  {\bibfield  {journal} {\bibinfo  {journal} {Nature}\ }\textbf {\bibinfo
  {volume} {546}},\ \bibinfo {pages} {129} (\bibinfo {year}
  {2017})}\BibitemShut {NoStop}%
\bibitem [{\citenamefont {Fukuzawa}\ \emph {et~al.}(2013)\citenamefont
  {Fukuzawa}, \citenamefont {Son}, \citenamefont {Motomura}, \citenamefont
  {Mondal}, \citenamefont {Nagaya}, \citenamefont {Wada}, \citenamefont {Liu},
  \citenamefont {Feifel}, \citenamefont {Tachibana}, \citenamefont {Ito},
  \citenamefont {Kimura}, \citenamefont {Sakai}, \citenamefont {Matsunami},
  \citenamefont {Hayashita}, \citenamefont {Kajikawa}, \citenamefont
  {Johnsson}, \citenamefont {Siano}, \citenamefont {Kukk}, \citenamefont
  {Rudek}, \citenamefont {Erk}, \citenamefont {Foucar}, \citenamefont {Robert},
  \citenamefont {Miron}, \citenamefont {Tono}, \citenamefont {Inubushi},
  \citenamefont {Hatsui}, \citenamefont {Yabashi}, \citenamefont {Yao},
  \citenamefont {Santra},\ and\ \citenamefont {Ueda}}]{Fukuzawa2012PRL}%
  \BibitemOpen
  \bibfield  {author} {\bibinfo {author} {\bibfnamefont {H.}~\bibnamefont
  {Fukuzawa}}, \bibinfo {author} {\bibfnamefont {S.-K.}\ \bibnamefont {Son}},
  \bibinfo {author} {\bibfnamefont {K.}~\bibnamefont {Motomura}}, \bibinfo
  {author} {\bibfnamefont {S.}~\bibnamefont {Mondal}}, \bibinfo {author}
  {\bibfnamefont {K.}~\bibnamefont {Nagaya}}, \bibinfo {author} {\bibfnamefont
  {S.}~\bibnamefont {Wada}}, \bibinfo {author} {\bibfnamefont {X.-J.}\
  \bibnamefont {Liu}}, \bibinfo {author} {\bibfnamefont {R.}~\bibnamefont
  {Feifel}}, \bibinfo {author} {\bibfnamefont {T.}~\bibnamefont {Tachibana}},
  \bibinfo {author} {\bibfnamefont {Y.}~\bibnamefont {Ito}}, \bibinfo {author}
  {\bibfnamefont {M.}~\bibnamefont {Kimura}}, \bibinfo {author} {\bibfnamefont
  {T.}~\bibnamefont {Sakai}}, \bibinfo {author} {\bibfnamefont
  {K.}~\bibnamefont {Matsunami}}, \bibinfo {author} {\bibfnamefont
  {H.}~\bibnamefont {Hayashita}}, \bibinfo {author} {\bibfnamefont
  {J.}~\bibnamefont {Kajikawa}}, \bibinfo {author} {\bibfnamefont
  {P.}~\bibnamefont {Johnsson}}, \bibinfo {author} {\bibfnamefont
  {M.}~\bibnamefont {Siano}}, \bibinfo {author} {\bibfnamefont
  {E.}~\bibnamefont {Kukk}}, \bibinfo {author} {\bibfnamefont {B.}~\bibnamefont
  {Rudek}}, \bibinfo {author} {\bibfnamefont {B.}~\bibnamefont {Erk}}, \bibinfo
  {author} {\bibfnamefont {L.}~\bibnamefont {Foucar}}, \bibinfo {author}
  {\bibfnamefont {E.}~\bibnamefont {Robert}}, \bibinfo {author} {\bibfnamefont
  {C.}~\bibnamefont {Miron}}, \bibinfo {author} {\bibfnamefont
  {K.}~\bibnamefont {Tono}}, \bibinfo {author} {\bibfnamefont {Y.}~\bibnamefont
  {Inubushi}}, \bibinfo {author} {\bibfnamefont {T.}~\bibnamefont {Hatsui}},
  \bibinfo {author} {\bibfnamefont {M.}~\bibnamefont {Yabashi}}, \bibinfo
  {author} {\bibfnamefont {M.}~\bibnamefont {Yao}}, \bibinfo {author}
  {\bibfnamefont {R.}~\bibnamefont {Santra}},\ and\ \bibinfo {author}
  {\bibfnamefont {K.}~\bibnamefont {Ueda}},\ }\bibfield  {title} {\bibinfo
  {title} {Deep inner-shell multiphoton ionization by intense x-ray
  free-electron laser pulses},\ }\href
  {https://doi.org/10.1103/PhysRevLett.110.173005} {\bibfield  {journal}
  {\bibinfo  {journal} {Phys. Rev. Lett.}\ }\textbf {\bibinfo {volume} {110}},\
  \bibinfo {pages} {173005} (\bibinfo {year} {2013})}\BibitemShut {NoStop}%
\bibitem [{\citenamefont {Doumy}\ \emph {et~al.}(2011)\citenamefont {Doumy},
  \citenamefont {Roedig}, \citenamefont {Son}, \citenamefont {Blaga},
  \citenamefont {DiChiara}, \citenamefont {Santra}, \citenamefont {Berrah},
  \citenamefont {Bostedt}, \citenamefont {Bozek}, \citenamefont {Bucksbaum}
  \emph {et~al.}}]{doumy2011nonlinear}%
  \BibitemOpen
  \bibfield  {author} {\bibinfo {author} {\bibfnamefont {G.}~\bibnamefont
  {Doumy}}, \bibinfo {author} {\bibfnamefont {C.}~\bibnamefont {Roedig}},
  \bibinfo {author} {\bibfnamefont {S.-K.}\ \bibnamefont {Son}}, \bibinfo
  {author} {\bibfnamefont {C.~I.}\ \bibnamefont {Blaga}}, \bibinfo {author}
  {\bibfnamefont {A.}~\bibnamefont {DiChiara}}, \bibinfo {author}
  {\bibfnamefont {R.}~\bibnamefont {Santra}}, \bibinfo {author} {\bibfnamefont
  {N.}~\bibnamefont {Berrah}}, \bibinfo {author} {\bibfnamefont
  {C.}~\bibnamefont {Bostedt}}, \bibinfo {author} {\bibfnamefont
  {J.}~\bibnamefont {Bozek}}, \bibinfo {author} {\bibfnamefont
  {P.}~\bibnamefont {Bucksbaum}}, \emph {et~al.},\ }\bibfield  {title}
  {\bibinfo {title} {Nonlinear atomic response to intense ultrashort x rays},\
  }\href@noop {} {\bibfield  {journal} {\bibinfo  {journal} {Physical Review
  Letters}\ }\textbf {\bibinfo {volume} {106}},\ \bibinfo {pages} {083002}
  (\bibinfo {year} {2011})}\BibitemShut {NoStop}%
\bibitem [{\citenamefont {Tamasaku}\ \emph {et~al.}(2014)\citenamefont
  {Tamasaku}, \citenamefont {Shigemasa}, \citenamefont {Inubushi},
  \citenamefont {Katayama}, \citenamefont {Sawada}, \citenamefont {Yumoto},
  \citenamefont {Ohashi}, \citenamefont {Mimura}, \citenamefont {Yabashi},
  \citenamefont {Yamauchi} \emph {et~al.}}]{tamasaku2014x}%
  \BibitemOpen
  \bibfield  {author} {\bibinfo {author} {\bibfnamefont {K.}~\bibnamefont
  {Tamasaku}}, \bibinfo {author} {\bibfnamefont {E.}~\bibnamefont {Shigemasa}},
  \bibinfo {author} {\bibfnamefont {Y.}~\bibnamefont {Inubushi}}, \bibinfo
  {author} {\bibfnamefont {T.}~\bibnamefont {Katayama}}, \bibinfo {author}
  {\bibfnamefont {K.}~\bibnamefont {Sawada}}, \bibinfo {author} {\bibfnamefont
  {H.}~\bibnamefont {Yumoto}}, \bibinfo {author} {\bibfnamefont
  {H.}~\bibnamefont {Ohashi}}, \bibinfo {author} {\bibfnamefont
  {H.}~\bibnamefont {Mimura}}, \bibinfo {author} {\bibfnamefont
  {M.}~\bibnamefont {Yabashi}}, \bibinfo {author} {\bibfnamefont
  {K.}~\bibnamefont {Yamauchi}}, \emph {et~al.},\ }\bibfield  {title} {\bibinfo
  {title} {X-ray two-photon absorption competing against single and sequential
  multiphoton processes},\ }\href@noop {} {\bibfield  {journal} {\bibinfo
  {journal} {Nature Photonics}\ }\textbf {\bibinfo {volume} {8}},\ \bibinfo
  {pages} {313} (\bibinfo {year} {2014})}\BibitemShut {NoStop}%
\bibitem [{\citenamefont {Ghimire}\ \emph {et~al.}(2016)\citenamefont
  {Ghimire}, \citenamefont {Fuchs}, \citenamefont {Hastings}, \citenamefont
  {Herrmann}, \citenamefont {Inubushi}, \citenamefont {Pines}, \citenamefont
  {Shwartz}, \citenamefont {Yabashi},\ and\ \citenamefont
  {Reis}}]{Ghimire2015PRA}%
  \BibitemOpen
  \bibfield  {author} {\bibinfo {author} {\bibfnamefont {S.}~\bibnamefont
  {Ghimire}}, \bibinfo {author} {\bibfnamefont {M.}~\bibnamefont {Fuchs}},
  \bibinfo {author} {\bibfnamefont {J.}~\bibnamefont {Hastings}}, \bibinfo
  {author} {\bibfnamefont {S.~C.}\ \bibnamefont {Herrmann}}, \bibinfo {author}
  {\bibfnamefont {Y.}~\bibnamefont {Inubushi}}, \bibinfo {author}
  {\bibfnamefont {J.}~\bibnamefont {Pines}}, \bibinfo {author} {\bibfnamefont
  {S.}~\bibnamefont {Shwartz}}, \bibinfo {author} {\bibfnamefont
  {M.}~\bibnamefont {Yabashi}},\ and\ \bibinfo {author} {\bibfnamefont {D.~A.}\
  \bibnamefont {Reis}},\ }\bibfield  {title} {\bibinfo {title} {Nonsequential
  two-photon absorption from the $k$ shell in solid zirconium},\ }\href
  {https://doi.org/10.1103/PhysRevA.94.043418} {\bibfield  {journal} {\bibinfo
  {journal} {Phys. Rev. A}\ }\textbf {\bibinfo {volume} {94}},\ \bibinfo
  {pages} {043418} (\bibinfo {year} {2016})}\BibitemShut {NoStop}%
\bibitem [{\citenamefont {Shwartz}\ \emph {et~al.}(2014)\citenamefont
  {Shwartz}, \citenamefont {Fuchs}, \citenamefont {Hastings}, \citenamefont
  {Inubushi}, \citenamefont {Ishikawa}, \citenamefont {Katayama}, \citenamefont
  {Reis}, \citenamefont {Sato}, \citenamefont {Tono}, \citenamefont {Yabashi}
  \emph {et~al.}}]{shwartz2014x}%
  \BibitemOpen
  \bibfield  {author} {\bibinfo {author} {\bibfnamefont {S.}~\bibnamefont
  {Shwartz}}, \bibinfo {author} {\bibfnamefont {M.}~\bibnamefont {Fuchs}},
  \bibinfo {author} {\bibfnamefont {J.}~\bibnamefont {Hastings}}, \bibinfo
  {author} {\bibfnamefont {Y.}~\bibnamefont {Inubushi}}, \bibinfo {author}
  {\bibfnamefont {T.}~\bibnamefont {Ishikawa}}, \bibinfo {author}
  {\bibfnamefont {T.}~\bibnamefont {Katayama}}, \bibinfo {author}
  {\bibfnamefont {D.}~\bibnamefont {Reis}}, \bibinfo {author} {\bibfnamefont
  {T.}~\bibnamefont {Sato}}, \bibinfo {author} {\bibfnamefont {K.}~\bibnamefont
  {Tono}}, \bibinfo {author} {\bibfnamefont {M.}~\bibnamefont {Yabashi}}, \emph
  {et~al.},\ }\bibfield  {title} {\bibinfo {title} {X-ray second harmonic
  generation},\ }\href@noop {} {\bibfield  {journal} {\bibinfo  {journal}
  {Physical review letters}\ }\textbf {\bibinfo {volume} {112}},\ \bibinfo
  {pages} {163901} (\bibinfo {year} {2014})}\BibitemShut {NoStop}%
\bibitem [{\citenamefont {Glover}\ \emph {et~al.}(2012)\citenamefont {Glover},
  \citenamefont {Fritz}, \citenamefont {Cammarata}, \citenamefont {Allison},
  \citenamefont {Coh}, \citenamefont {Feldkamp}, \citenamefont {Lemke},
  \citenamefont {Zhu}, \citenamefont {Feng}, \citenamefont {Coffee} \emph
  {et~al.}}]{glover2012x}%
  \BibitemOpen
  \bibfield  {author} {\bibinfo {author} {\bibfnamefont {T.}~\bibnamefont
  {Glover}}, \bibinfo {author} {\bibfnamefont {D.}~\bibnamefont {Fritz}},
  \bibinfo {author} {\bibfnamefont {M.}~\bibnamefont {Cammarata}}, \bibinfo
  {author} {\bibfnamefont {T.}~\bibnamefont {Allison}}, \bibinfo {author}
  {\bibfnamefont {S.}~\bibnamefont {Coh}}, \bibinfo {author} {\bibfnamefont
  {J.}~\bibnamefont {Feldkamp}}, \bibinfo {author} {\bibfnamefont
  {H.}~\bibnamefont {Lemke}}, \bibinfo {author} {\bibfnamefont
  {D.}~\bibnamefont {Zhu}}, \bibinfo {author} {\bibfnamefont {Y.}~\bibnamefont
  {Feng}}, \bibinfo {author} {\bibfnamefont {R.}~\bibnamefont {Coffee}}, \emph
  {et~al.},\ }\bibfield  {title} {\bibinfo {title} {X-ray and optical wave
  mixing},\ }\href@noop {} {\bibfield  {journal} {\bibinfo  {journal} {Nature}\
  }\textbf {\bibinfo {volume} {488}},\ \bibinfo {pages} {603} (\bibinfo {year}
  {2012})}\BibitemShut {NoStop}%
\bibitem [{\citenamefont {Rohringer}\ \emph {et~al.}(2012)\citenamefont
  {Rohringer}, \citenamefont {Ryan}, \citenamefont {London}, \citenamefont
  {Purvis}, \citenamefont {Albert}, \citenamefont {Dunn}, \citenamefont
  {Bozek}, \citenamefont {Bostedt}, \citenamefont {Graf}, \citenamefont {Hill}
  \emph {et~al.}}]{rohringer2012atomic}%
  \BibitemOpen
  \bibfield  {author} {\bibinfo {author} {\bibfnamefont {N.}~\bibnamefont
  {Rohringer}}, \bibinfo {author} {\bibfnamefont {D.}~\bibnamefont {Ryan}},
  \bibinfo {author} {\bibfnamefont {R.~A.}\ \bibnamefont {London}}, \bibinfo
  {author} {\bibfnamefont {M.}~\bibnamefont {Purvis}}, \bibinfo {author}
  {\bibfnamefont {F.}~\bibnamefont {Albert}}, \bibinfo {author} {\bibfnamefont
  {J.}~\bibnamefont {Dunn}}, \bibinfo {author} {\bibfnamefont {J.~D.}\
  \bibnamefont {Bozek}}, \bibinfo {author} {\bibfnamefont {C.}~\bibnamefont
  {Bostedt}}, \bibinfo {author} {\bibfnamefont {A.}~\bibnamefont {Graf}},
  \bibinfo {author} {\bibfnamefont {R.}~\bibnamefont {Hill}}, \emph {et~al.},\
  }\bibfield  {title} {\bibinfo {title} {Atomic inner-shell x-ray laser at 1.46
  nanometres pumped by an x-ray free-electron laser},\ }\href@noop {}
  {\bibfield  {journal} {\bibinfo  {journal} {Nature}\ }\textbf {\bibinfo
  {volume} {481}},\ \bibinfo {pages} {488} (\bibinfo {year}
  {2012})}\BibitemShut {NoStop}%
\bibitem [{\citenamefont {Beye}\ \emph {et~al.}(2013)\citenamefont {Beye},
  \citenamefont {Schreck}, \citenamefont {Sorgenfrei}, \citenamefont {Trabant},
  \citenamefont {Pontius}, \citenamefont {Sch{\"u}{\ss}ler-Langeheine},
  \citenamefont {Wurth},\ and\ \citenamefont
  {F{\"o}hlisch}}]{beye2013stimulated}%
  \BibitemOpen
  \bibfield  {author} {\bibinfo {author} {\bibfnamefont {M.}~\bibnamefont
  {Beye}}, \bibinfo {author} {\bibfnamefont {S.}~\bibnamefont {Schreck}},
  \bibinfo {author} {\bibfnamefont {F.}~\bibnamefont {Sorgenfrei}}, \bibinfo
  {author} {\bibfnamefont {C.}~\bibnamefont {Trabant}}, \bibinfo {author}
  {\bibfnamefont {N.}~\bibnamefont {Pontius}}, \bibinfo {author} {\bibfnamefont
  {C.}~\bibnamefont {Sch{\"u}{\ss}ler-Langeheine}}, \bibinfo {author}
  {\bibfnamefont {W.}~\bibnamefont {Wurth}},\ and\ \bibinfo {author}
  {\bibfnamefont {A.}~\bibnamefont {F{\"o}hlisch}},\ }\bibfield  {title}
  {\bibinfo {title} {Stimulated x-ray emission for materials science},\
  }\href@noop {} {\bibfield  {journal} {\bibinfo  {journal} {Nature}\ }\textbf
  {\bibinfo {volume} {501}},\ \bibinfo {pages} {191} (\bibinfo {year}
  {2013})}\BibitemShut {NoStop}%
\bibitem [{\citenamefont {Weninger}\ \emph {et~al.}(2013)\citenamefont
  {Weninger}, \citenamefont {Purvis}, \citenamefont {Ryan}, \citenamefont
  {London}, \citenamefont {Bozek}, \citenamefont {Bostedt}, \citenamefont
  {Graf}, \citenamefont {Brown}, \citenamefont {Rocca},\ and\ \citenamefont
  {Rohringer}}]{weninger2013stimulatedPRL}%
  \BibitemOpen
  \bibfield  {author} {\bibinfo {author} {\bibfnamefont {C.}~\bibnamefont
  {Weninger}}, \bibinfo {author} {\bibfnamefont {M.}~\bibnamefont {Purvis}},
  \bibinfo {author} {\bibfnamefont {D.}~\bibnamefont {Ryan}}, \bibinfo {author}
  {\bibfnamefont {R.~A.}\ \bibnamefont {London}}, \bibinfo {author}
  {\bibfnamefont {J.~D.}\ \bibnamefont {Bozek}}, \bibinfo {author}
  {\bibfnamefont {C.}~\bibnamefont {Bostedt}}, \bibinfo {author} {\bibfnamefont
  {A.}~\bibnamefont {Graf}}, \bibinfo {author} {\bibfnamefont {G.}~\bibnamefont
  {Brown}}, \bibinfo {author} {\bibfnamefont {J.~J.}\ \bibnamefont {Rocca}},\
  and\ \bibinfo {author} {\bibfnamefont {N.}~\bibnamefont {Rohringer}},\
  }\bibfield  {title} {\bibinfo {title} {Stimulated electronic x-ray raman
  scattering},\ }\href@noop {} {\bibfield  {journal} {\bibinfo  {journal}
  {Physical review letters}\ }\textbf {\bibinfo {volume} {111}},\ \bibinfo
  {pages} {233902} (\bibinfo {year} {2013})}\BibitemShut {NoStop}%
\bibitem [{\citenamefont {Yoneda}\ \emph {et~al.}(2015)\citenamefont {Yoneda},
  \citenamefont {Inubushi}, \citenamefont {Nagamine}, \citenamefont {Michine},
  \citenamefont {Ohashi}, \citenamefont {Yumoto}, \citenamefont {Yamauchi},
  \citenamefont {Mimura}, \citenamefont {Kitamura}, \citenamefont {Katayama},
  \citenamefont {Ishikawa},\ and\ \citenamefont {Yabashi}}]{Yoneda2015Nat}%
  \BibitemOpen
  \bibfield  {author} {\bibinfo {author} {\bibfnamefont {H.}~\bibnamefont
  {Yoneda}}, \bibinfo {author} {\bibfnamefont {Y.}~\bibnamefont {Inubushi}},
  \bibinfo {author} {\bibfnamefont {K.}~\bibnamefont {Nagamine}}, \bibinfo
  {author} {\bibfnamefont {Y.}~\bibnamefont {Michine}}, \bibinfo {author}
  {\bibfnamefont {H.}~\bibnamefont {Ohashi}}, \bibinfo {author} {\bibfnamefont
  {H.}~\bibnamefont {Yumoto}}, \bibinfo {author} {\bibfnamefont
  {K.}~\bibnamefont {Yamauchi}}, \bibinfo {author} {\bibfnamefont
  {H.}~\bibnamefont {Mimura}}, \bibinfo {author} {\bibfnamefont
  {H.}~\bibnamefont {Kitamura}}, \bibinfo {author} {\bibfnamefont
  {T.}~\bibnamefont {Katayama}}, \bibinfo {author} {\bibfnamefont
  {T.}~\bibnamefont {Ishikawa}},\ and\ \bibinfo {author} {\bibfnamefont
  {M.}~\bibnamefont {Yabashi}},\ }\bibfield  {title} {\bibinfo {title} {Atomic
  inner-shell laser at 1.5-{\aa}ngstr{\"o}m wavelength pumped by an x-ray
  free-electron laser},\ }\href {https://doi.org/10.1038/nature14894}
  {\bibfield  {journal} {\bibinfo  {journal} {Nature}\ }\textbf {\bibinfo
  {volume} {524}},\ \bibinfo {pages} {446} (\bibinfo {year}
  {2015})}\BibitemShut {NoStop}%
\bibitem [{\citenamefont {Bloch}(1946)}]{BlochPR1946}%
  \BibitemOpen
  \bibfield  {author} {\bibinfo {author} {\bibfnamefont {F.}~\bibnamefont
  {Bloch}},\ }\bibfield  {title} {\bibinfo {title} {Nuclear induction},\ }\href
  {https://doi.org/10.1103/PhysRev.70.460} {\bibfield  {journal} {\bibinfo
  {journal} {Phys. Rev.}\ }\textbf {\bibinfo {volume} {70}},\ \bibinfo {pages}
  {460} (\bibinfo {year} {1946})}\BibitemShut {NoStop}%
\bibitem [{\citenamefont {Brewer}\ and\ \citenamefont
  {Shoemaker}(1972)}]{brewerPRA1972}%
  \BibitemOpen
  \bibfield  {author} {\bibinfo {author} {\bibfnamefont {R.~G.}\ \bibnamefont
  {Brewer}}\ and\ \bibinfo {author} {\bibfnamefont {R.~L.}\ \bibnamefont
  {Shoemaker}},\ }\bibfield  {title} {\bibinfo {title} {Optical free induction
  decay},\ }\href {https://doi.org/10.1103/PhysRevA.6.2001} {\bibfield
  {journal} {\bibinfo  {journal} {Phys. Rev. A}\ }\textbf {\bibinfo {volume}
  {6}},\ \bibinfo {pages} {2001} (\bibinfo {year} {1972})}\BibitemShut
  {NoStop}%
\bibitem [{\citenamefont {McCall}\ and\ \citenamefont
  {Hahn}(1969)}]{mccall1969self}%
  \BibitemOpen
  \bibfield  {author} {\bibinfo {author} {\bibfnamefont {S.~L.}\ \bibnamefont
  {McCall}}\ and\ \bibinfo {author} {\bibfnamefont {E.~L.}\ \bibnamefont
  {Hahn}},\ }\bibfield  {title} {\bibinfo {title} {Self-induced transparency},\
  }\href@noop {} {\bibfield  {journal} {\bibinfo  {journal} {Physical Review}\
  }\textbf {\bibinfo {volume} {183}},\ \bibinfo {pages} {457} (\bibinfo {year}
  {1969})}\BibitemShut {NoStop}%
\bibitem [{\citenamefont {McCall}\ and\ \citenamefont
  {Hahn}(1967)}]{mccall1967self}%
  \BibitemOpen
  \bibfield  {author} {\bibinfo {author} {\bibfnamefont {S.~L.}\ \bibnamefont
  {McCall}}\ and\ \bibinfo {author} {\bibfnamefont {E.~L.}\ \bibnamefont
  {Hahn}},\ }\bibfield  {title} {\bibinfo {title} {Self-induced transparency by
  pulsed coherent light},\ }\href@noop {} {\bibfield  {journal} {\bibinfo
  {journal} {Physical Review Letters}\ }\textbf {\bibinfo {volume} {18}},\
  \bibinfo {pages} {908} (\bibinfo {year} {1967})}\BibitemShut {NoStop}%
\bibitem [{\citenamefont {Crisp}(1970)}]{crisp1970propagation}%
  \BibitemOpen
  \bibfield  {author} {\bibinfo {author} {\bibfnamefont {M.}~\bibnamefont
  {Crisp}},\ }\bibfield  {title} {\bibinfo {title} {Propagation of small-area
  pulses of coherent light through a resonant medium},\ }\href@noop {}
  {\bibfield  {journal} {\bibinfo  {journal} {Physical Review A}\ }\textbf
  {\bibinfo {volume} {1}},\ \bibinfo {pages} {1604} (\bibinfo {year}
  {1970})}\BibitemShut {NoStop}%
\bibitem [{\citenamefont {Chiao}\ \emph {et~al.}(1964)\citenamefont {Chiao},
  \citenamefont {Garmire},\ and\ \citenamefont {Townes}}]{chiao1964self}%
  \BibitemOpen
  \bibfield  {author} {\bibinfo {author} {\bibfnamefont {R.~Y.}\ \bibnamefont
  {Chiao}}, \bibinfo {author} {\bibfnamefont {E.}~\bibnamefont {Garmire}},\
  and\ \bibinfo {author} {\bibfnamefont {C.~H.}\ \bibnamefont {Townes}},\
  }\bibfield  {title} {\bibinfo {title} {Self-trapping of optical beams},\
  }\href@noop {} {\bibfield  {journal} {\bibinfo  {journal} {Physical Review
  Letters}\ }\textbf {\bibinfo {volume} {13}},\ \bibinfo {pages} {479}
  (\bibinfo {year} {1964})}\BibitemShut {NoStop}%
\bibitem [{\citenamefont {Gibbs}\ \emph {et~al.}(1976)\citenamefont {Gibbs},
  \citenamefont {B{\"o}lger}, \citenamefont {Mattar}, \citenamefont {Newstein},
  \citenamefont {Forster},\ and\ \citenamefont {Toschek}}]{gibbs1976coherent}%
  \BibitemOpen
  \bibfield  {author} {\bibinfo {author} {\bibfnamefont {H.}~\bibnamefont
  {Gibbs}}, \bibinfo {author} {\bibfnamefont {B.}~\bibnamefont {B{\"o}lger}},
  \bibinfo {author} {\bibfnamefont {F.}~\bibnamefont {Mattar}}, \bibinfo
  {author} {\bibfnamefont {M.}~\bibnamefont {Newstein}}, \bibinfo {author}
  {\bibfnamefont {G.}~\bibnamefont {Forster}},\ and\ \bibinfo {author}
  {\bibfnamefont {P.}~\bibnamefont {Toschek}},\ }\bibfield  {title} {\bibinfo
  {title} {Coherent on-resonance self-focusing of optical pulses in
  absorbers},\ }\href@noop {} {\bibfield  {journal} {\bibinfo  {journal}
  {Physical Review Letters}\ }\textbf {\bibinfo {volume} {37}},\ \bibinfo
  {pages} {1743} (\bibinfo {year} {1976})}\BibitemShut {NoStop}%
\bibitem [{\citenamefont {Wright}\ and\ \citenamefont
  {Newstein}(1973)}]{wright1973self}%
  \BibitemOpen
  \bibfield  {author} {\bibinfo {author} {\bibfnamefont {N.}~\bibnamefont
  {Wright}}\ and\ \bibinfo {author} {\bibfnamefont {M.~C.}\ \bibnamefont
  {Newstein}},\ }\bibfield  {title} {\bibinfo {title} {Self-focusing of
  coherent pulses},\ }\href@noop {} {\bibfield  {journal} {\bibinfo  {journal}
  {Optics Communications}\ }\textbf {\bibinfo {volume} {9}},\ \bibinfo {pages}
  {8} (\bibinfo {year} {1973})}\BibitemShut {NoStop}%
\bibitem [{\citenamefont {Burnham}\ and\ \citenamefont
  {Chiao}(1969)}]{Burnham1969PR}%
  \BibitemOpen
  \bibfield  {author} {\bibinfo {author} {\bibfnamefont {D.~C.}\ \bibnamefont
  {Burnham}}\ and\ \bibinfo {author} {\bibfnamefont {R.~Y.}\ \bibnamefont
  {Chiao}},\ }\bibfield  {title} {\bibinfo {title} {Coherent resonance
  fluorescence excited by short light pulses},\ }\href
  {https://doi.org/10.1103/PhysRev.188.667} {\bibfield  {journal} {\bibinfo
  {journal} {Phys. Rev.}\ }\textbf {\bibinfo {volume} {188}},\ \bibinfo {pages}
  {667} (\bibinfo {year} {1969})}\BibitemShut {NoStop}%
\bibitem [{\citenamefont {Pfeiffer}\ \emph {et~al.}(2013)\citenamefont
  {Pfeiffer}, \citenamefont {Bell}, \citenamefont {Beck}, \citenamefont
  {Mashiko}, \citenamefont {Neumark},\ and\ \citenamefont
  {Leone}}]{Pfeiffer2013.MacroscopicAbsorption}%
  \BibitemOpen
  \bibfield  {author} {\bibinfo {author} {\bibfnamefont {A.~A.}\ \bibnamefont
  {Pfeiffer}}, \bibinfo {author} {\bibfnamefont {J.~B.}\ \bibnamefont {Bell}},
  \bibinfo {author} {\bibfnamefont {A.~R.}\ \bibnamefont {Beck}}, \bibinfo
  {author} {\bibfnamefont {H.}~\bibnamefont {Mashiko}}, \bibinfo {author}
  {\bibfnamefont {D.~M.}\ \bibnamefont {Neumark}},\ and\ \bibinfo {author}
  {\bibfnamefont {S.~R.~L.}\ \bibnamefont {Leone}},\ }\bibfield  {title}
  {\bibinfo {title} {Alternating absorption features during attosecond-pulse
  propagation in a laser-controlled gaseous medium},\ }\href@noop {} {\bibfield
   {journal} {\bibinfo  {journal} {Physical Review A}\ }\textbf {\bibinfo
  {volume} {88}},\ \bibinfo {pages} {051402(R)} (\bibinfo {year}
  {2013})}\BibitemShut {NoStop}%
\bibitem [{\citenamefont {Liao}\ \emph {et~al.}(2015)\citenamefont {Liao},
  \citenamefont {Sandhu}, \citenamefont {Camp}, \citenamefont {Schafer},\ and\
  \citenamefont {Gaarde}}]{liao2015beyond}%
  \BibitemOpen
  \bibfield  {author} {\bibinfo {author} {\bibfnamefont {C.-T.}\ \bibnamefont
  {Liao}}, \bibinfo {author} {\bibfnamefont {A.}~\bibnamefont {Sandhu}},
  \bibinfo {author} {\bibfnamefont {S.}~\bibnamefont {Camp}}, \bibinfo {author}
  {\bibfnamefont {K.~J.}\ \bibnamefont {Schafer}},\ and\ \bibinfo {author}
  {\bibfnamefont {M.~B.}\ \bibnamefont {Gaarde}},\ }\bibfield  {title}
  {\bibinfo {title} {Beyond the single-atom response in absorption line shapes:
  probing a dense, laser-dressed helium gas with attosecond pulse trains},\
  }\href@noop {} {\bibfield  {journal} {\bibinfo  {journal} {Physical review
  letters}\ }\textbf {\bibinfo {volume} {114}},\ \bibinfo {pages} {143002}
  (\bibinfo {year} {2015})}\BibitemShut {NoStop}%
\bibitem [{\citenamefont {Bengtsson}\ \emph {et~al.}(2017)\citenamefont
  {Bengtsson}, \citenamefont {Larsen}, \citenamefont {Kroon}, \citenamefont
  {Camp}, \citenamefont {Miranda}, \citenamefont {Arnold}, \citenamefont
  {L'Huillier}, \citenamefont {Schafer}, \citenamefont {Gaarde}, \citenamefont
  {Rippe},\ and\ \citenamefont {Mauritsson}}]{bengtsson_spacetime_2017}%
  \BibitemOpen
  \bibfield  {author} {\bibinfo {author} {\bibfnamefont {S.}~\bibnamefont
  {Bengtsson}}, \bibinfo {author} {\bibfnamefont {E.~W.}\ \bibnamefont
  {Larsen}}, \bibinfo {author} {\bibfnamefont {D.}~\bibnamefont {Kroon}},
  \bibinfo {author} {\bibfnamefont {S.}~\bibnamefont {Camp}}, \bibinfo {author}
  {\bibfnamefont {M.}~\bibnamefont {Miranda}}, \bibinfo {author} {\bibfnamefont
  {C.~L.}\ \bibnamefont {Arnold}}, \bibinfo {author} {\bibfnamefont
  {A.}~\bibnamefont {L'Huillier}}, \bibinfo {author} {\bibfnamefont {K.~J.}\
  \bibnamefont {Schafer}}, \bibinfo {author} {\bibfnamefont {M.~B.}\
  \bibnamefont {Gaarde}}, \bibinfo {author} {\bibfnamefont {L.}~\bibnamefont
  {Rippe}},\ and\ \bibinfo {author} {\bibfnamefont {J.}~\bibnamefont
  {Mauritsson}},\ }\bibfield  {title} {\bibinfo {title} {Space–time control
  of free induction decay in the extreme ultraviolet},\ }\href
  {https://doi.org/10.1038/nphoton.2017.30} {\bibfield  {journal} {\bibinfo
  {journal} {Nature Photonics}\ }\textbf {\bibinfo {volume} {11}},\ \bibinfo
  {pages} {252} (\bibinfo {year} {2017})}\BibitemShut {NoStop}%
\bibitem [{\citenamefont {Harries}\ \emph {et~al.}(2018)\citenamefont
  {Harries}, \citenamefont {Iwayama}, \citenamefont {Kuma}, \citenamefont
  {Iizawa}, \citenamefont {Suzuki}, \citenamefont {Azuma}, \citenamefont
  {Inoue}, \citenamefont {Owada}, \citenamefont {Togashi}, \citenamefont
  {Tono}, \citenamefont {Yabashi},\ and\ \citenamefont
  {Shigemasa}}]{Harries2018PRL}%
  \BibitemOpen
  \bibfield  {author} {\bibinfo {author} {\bibfnamefont {J.~R.}\ \bibnamefont
  {Harries}}, \bibinfo {author} {\bibfnamefont {H.}~\bibnamefont {Iwayama}},
  \bibinfo {author} {\bibfnamefont {S.}~\bibnamefont {Kuma}}, \bibinfo {author}
  {\bibfnamefont {M.}~\bibnamefont {Iizawa}}, \bibinfo {author} {\bibfnamefont
  {N.}~\bibnamefont {Suzuki}}, \bibinfo {author} {\bibfnamefont
  {Y.}~\bibnamefont {Azuma}}, \bibinfo {author} {\bibfnamefont
  {I.}~\bibnamefont {Inoue}}, \bibinfo {author} {\bibfnamefont
  {S.}~\bibnamefont {Owada}}, \bibinfo {author} {\bibfnamefont
  {T.}~\bibnamefont {Togashi}}, \bibinfo {author} {\bibfnamefont
  {K.}~\bibnamefont {Tono}}, \bibinfo {author} {\bibfnamefont {M.}~\bibnamefont
  {Yabashi}},\ and\ \bibinfo {author} {\bibfnamefont {E.}~\bibnamefont
  {Shigemasa}},\ }\bibfield  {title} {\bibinfo {title} {Superfluorescence,
  free-induction decay, and four-wave mixing: Propagation of free-electron
  laser pulses through a dense sample of helium ions},\ }\href
  {https://doi.org/10.1103/PhysRevLett.121.263201} {\bibfield  {journal}
  {\bibinfo  {journal} {Phys. Rev. Lett.}\ }\textbf {\bibinfo {volume} {121}},\
  \bibinfo {pages} {263201} (\bibinfo {year} {2018})}\BibitemShut {NoStop}%
\bibitem [{\citenamefont {Marinelli}\ \emph {et~al.}(2017)\citenamefont
  {Marinelli}, \citenamefont {MacArthur}, \citenamefont {Emma}, \citenamefont
  {Guetg}, \citenamefont {Field}, \citenamefont {Kharakh}, \citenamefont
  {Lutman}, \citenamefont {Ding},\ and\ \citenamefont
  {Huang}}]{marinelli2017experimental}%
  \BibitemOpen
  \bibfield  {author} {\bibinfo {author} {\bibfnamefont {A.}~\bibnamefont
  {Marinelli}}, \bibinfo {author} {\bibfnamefont {J.}~\bibnamefont
  {MacArthur}}, \bibinfo {author} {\bibfnamefont {P.}~\bibnamefont {Emma}},
  \bibinfo {author} {\bibfnamefont {M.}~\bibnamefont {Guetg}}, \bibinfo
  {author} {\bibfnamefont {C.}~\bibnamefont {Field}}, \bibinfo {author}
  {\bibfnamefont {D.}~\bibnamefont {Kharakh}}, \bibinfo {author} {\bibfnamefont
  {A.}~\bibnamefont {Lutman}}, \bibinfo {author} {\bibfnamefont
  {Y.}~\bibnamefont {Ding}},\ and\ \bibinfo {author} {\bibfnamefont
  {Z.}~\bibnamefont {Huang}},\ }\bibfield  {title} {\bibinfo {title}
  {Experimental demonstration of a single-spike hard-x-ray free-electron laser
  starting from noise},\ }\href@noop {} {\bibfield  {journal} {\bibinfo
  {journal} {Applied Physics Letters}\ }\textbf {\bibinfo {volume} {111}},\
  \bibinfo {pages} {151101} (\bibinfo {year} {2017})}\BibitemShut {NoStop}%
\bibitem [{\citenamefont {Huang}\ \emph {et~al.}(2017)\citenamefont {Huang},
  \citenamefont {Ding}, \citenamefont {Feng}, \citenamefont {Hemsing},
  \citenamefont {Huang}, \citenamefont {Krzywinski}, \citenamefont {Lutman},
  \citenamefont {Marinelli}, \citenamefont {Maxwell},\ and\ \citenamefont
  {Zhu}}]{huang2017generating}%
  \BibitemOpen
  \bibfield  {author} {\bibinfo {author} {\bibfnamefont {S.}~\bibnamefont
  {Huang}}, \bibinfo {author} {\bibfnamefont {Y.}~\bibnamefont {Ding}},
  \bibinfo {author} {\bibfnamefont {Y.}~\bibnamefont {Feng}}, \bibinfo {author}
  {\bibfnamefont {E.}~\bibnamefont {Hemsing}}, \bibinfo {author} {\bibfnamefont
  {Z.}~\bibnamefont {Huang}}, \bibinfo {author} {\bibfnamefont
  {J.}~\bibnamefont {Krzywinski}}, \bibinfo {author} {\bibfnamefont
  {A.}~\bibnamefont {Lutman}}, \bibinfo {author} {\bibfnamefont
  {A.}~\bibnamefont {Marinelli}}, \bibinfo {author} {\bibfnamefont
  {T.}~\bibnamefont {Maxwell}},\ and\ \bibinfo {author} {\bibfnamefont
  {D.}~\bibnamefont {Zhu}},\ }\bibfield  {title} {\bibinfo {title} {Generating
  single-spike hard x-ray pulses with nonlinear bunch compression in
  free-electron lasers},\ }\href@noop {} {\bibfield  {journal} {\bibinfo
  {journal} {Physical review letters}\ }\textbf {\bibinfo {volume} {119}},\
  \bibinfo {pages} {154801} (\bibinfo {year} {2017})}\BibitemShut {NoStop}%
\bibitem [{\citenamefont {Duris}\ \emph {et~al.}(2020)\citenamefont {Duris},
  \citenamefont {Li}, \citenamefont {Driver}, \citenamefont {Champenois},
  \citenamefont {MacArthur}, \citenamefont {Lutman}, \citenamefont {Zhang},
  \citenamefont {Rosenberger}, \citenamefont {Aldrich}, \citenamefont {Coffee}
  \emph {et~al.}}]{duris2020tunable}%
  \BibitemOpen
  \bibfield  {author} {\bibinfo {author} {\bibfnamefont {J.}~\bibnamefont
  {Duris}}, \bibinfo {author} {\bibfnamefont {S.}~\bibnamefont {Li}}, \bibinfo
  {author} {\bibfnamefont {T.}~\bibnamefont {Driver}}, \bibinfo {author}
  {\bibfnamefont {E.~G.}\ \bibnamefont {Champenois}}, \bibinfo {author}
  {\bibfnamefont {J.~P.}\ \bibnamefont {MacArthur}}, \bibinfo {author}
  {\bibfnamefont {A.~A.}\ \bibnamefont {Lutman}}, \bibinfo {author}
  {\bibfnamefont {Z.}~\bibnamefont {Zhang}}, \bibinfo {author} {\bibfnamefont
  {P.}~\bibnamefont {Rosenberger}}, \bibinfo {author} {\bibfnamefont {J.~W.}\
  \bibnamefont {Aldrich}}, \bibinfo {author} {\bibfnamefont {R.}~\bibnamefont
  {Coffee}}, \emph {et~al.},\ }\bibfield  {title} {\bibinfo {title} {Tunable
  isolated attosecond x-ray pulses with gigawatt peak power from a
  free-electron laser},\ }\href@noop {} {\bibfield  {journal} {\bibinfo
  {journal} {Nature Photonics}\ }\textbf {\bibinfo {volume} {14}},\ \bibinfo
  {pages} {30} (\bibinfo {year} {2020})}\BibitemShut {NoStop}%
\bibitem [{\citenamefont {Sun}\ \emph {et~al.}(2010)\citenamefont {Sun},
  \citenamefont {Liu}, \citenamefont {Wang},\ and\ \citenamefont
  {Gel’mukhanov}}]{sun2010propagation}%
  \BibitemOpen
  \bibfield  {author} {\bibinfo {author} {\bibfnamefont {Y.-P.}\ \bibnamefont
  {Sun}}, \bibinfo {author} {\bibfnamefont {J.-C.}\ \bibnamefont {Liu}},
  \bibinfo {author} {\bibfnamefont {C.-K.}\ \bibnamefont {Wang}},\ and\
  \bibinfo {author} {\bibfnamefont {F.}~\bibnamefont {Gel’mukhanov}},\
  }\bibfield  {title} {\bibinfo {title} {Propagation of a strong x-ray pulse:
  Pulse compression, stimulated raman scattering, amplified spontaneous
  emission, lasing without inversion, and four-wave mixing},\ }\href@noop {}
  {\bibfield  {journal} {\bibinfo  {journal} {Physical Review A}\ }\textbf
  {\bibinfo {volume} {81}},\ \bibinfo {pages} {013812} (\bibinfo {year}
  {2010})}\BibitemShut {NoStop}%
\bibitem [{\citenamefont {Weninger}\ and\ \citenamefont
  {Rohringer}(2013)}]{weninger2013PRA}%
  \BibitemOpen
  \bibfield  {author} {\bibinfo {author} {\bibfnamefont {C.}~\bibnamefont
  {Weninger}}\ and\ \bibinfo {author} {\bibfnamefont {N.}~\bibnamefont
  {Rohringer}},\ }\bibfield  {title} {\bibinfo {title} {Stimulated resonant
  x-ray raman scattering with incoherent radiation},\ }\href@noop {} {\bibfield
   {journal} {\bibinfo  {journal} {Physical Review A}\ }\textbf {\bibinfo
  {volume} {88}},\ \bibinfo {pages} {053421} (\bibinfo {year}
  {2013})}\BibitemShut {NoStop}%
\bibitem [{\citenamefont {Lyu}\ \emph {et~al.}(2020)\citenamefont {Lyu},
  \citenamefont {Cavaletto}, \citenamefont {Keitel},\ and\ \citenamefont
  {Harman}}]{lyu2020narrow}%
  \BibitemOpen
  \bibfield  {author} {\bibinfo {author} {\bibfnamefont {C.}~\bibnamefont
  {Lyu}}, \bibinfo {author} {\bibfnamefont {S.~M.}\ \bibnamefont {Cavaletto}},
  \bibinfo {author} {\bibfnamefont {C.~H.}\ \bibnamefont {Keitel}},\ and\
  \bibinfo {author} {\bibfnamefont {Z.}~\bibnamefont {Harman}},\ }\bibfield
  {title} {\bibinfo {title} {Narrow-band hard-x-ray lasing with highly charged
  ions},\ }\href@noop {} {\bibfield  {journal} {\bibinfo  {journal} {Scientific
  Reports}\ }\textbf {\bibinfo {volume} {10}},\ \bibinfo {pages} {1} (\bibinfo
  {year} {2020})}\BibitemShut {NoStop}%
\bibitem [{\citenamefont {Cowan}(1981)}]{Cowan}%
  \BibitemOpen
  \bibfield  {author} {\bibinfo {author} {\bibfnamefont {R.~D.}\ \bibnamefont
  {Cowan}},\ }\href@noop {} {\emph {\bibinfo {title} {The Theory of Atomic
  Structure and Spectra}}}\ (\bibinfo  {publisher} {University of California
  Press, Berkeley},\ \bibinfo {year} {1981})\BibitemShut {NoStop}%
\bibitem [{\citenamefont {Breinig}\ \emph {et~al.}(1980)\citenamefont
  {Breinig}, \citenamefont {Chen}, \citenamefont {Ice}, \citenamefont
  {Parente}, \citenamefont {Crasemann},\ and\ \citenamefont
  {Brown}}]{breinig_atomic_1980}%
  \BibitemOpen
  \bibfield  {author} {\bibinfo {author} {\bibfnamefont {M.}~\bibnamefont
  {Breinig}}, \bibinfo {author} {\bibfnamefont {M.~H.}\ \bibnamefont {Chen}},
  \bibinfo {author} {\bibfnamefont {G.~E.}\ \bibnamefont {Ice}}, \bibinfo
  {author} {\bibfnamefont {F.}~\bibnamefont {Parente}}, \bibinfo {author}
  {\bibfnamefont {B.}~\bibnamefont {Crasemann}},\ and\ \bibinfo {author}
  {\bibfnamefont {G.~S.}\ \bibnamefont {Brown}},\ }\bibfield  {title} {\bibinfo
  {title} {Atomic inner-shell level energies determined by absorption
  spectrometry with synchrotron radiation},\ }\href
  {https://doi.org/10.1103/PhysRevA.22.520} {\bibfield  {journal} {\bibinfo
  {journal} {Physical Review A}\ }\textbf {\bibinfo {volume} {22}},\ \bibinfo
  {pages} {520} (\bibinfo {year} {1980})}\BibitemShut {NoStop}%
\bibitem [{\citenamefont {Frigo}\ and\ \citenamefont
  {Johnson}(2005)}]{fftw_article}%
  \BibitemOpen
  \bibfield  {author} {\bibinfo {author} {\bibfnamefont {M.}~\bibnamefont
  {Frigo}}\ and\ \bibinfo {author} {\bibfnamefont {S.~G.}\ \bibnamefont
  {Johnson}},\ }\bibfield  {title} {\bibinfo {title} {The {Design} and
  {Implementation} of {FFTW}3},\ }\href
  {https://doi.org/10.1109/JPROC.2004.840301} {\bibfield  {journal} {\bibinfo
  {journal} {Proceedings of the IEEE}\ }\textbf {\bibinfo {volume} {93}},\
  \bibinfo {pages} {216} (\bibinfo {year} {2005})}\BibitemShut {NoStop}%
\bibitem [{lap()}]{lapack_website}%
  \BibitemOpen
  \href@noop {} {\bibinfo {title} {Linear algebra package}},\ \bibinfo
  {howpublished} {\url{http://http://www.netlib.org/lapack}}\BibitemShut
  {NoStop}%
\bibitem [{\citenamefont {Slusher}\ and\ \citenamefont
  {Gibbs}(1972)}]{slusher1972PRA}%
  \BibitemOpen
  \bibfield  {author} {\bibinfo {author} {\bibfnamefont {R.~E.}\ \bibnamefont
  {Slusher}}\ and\ \bibinfo {author} {\bibfnamefont {H.~M.}\ \bibnamefont
  {Gibbs}},\ }\bibfield  {title} {\bibinfo {title} {Self-induced transparency
  in atomic rubidium,},\ }\href {https://doi.org/10.1103/PhysRevA.6.1255.3}
  {\bibfield  {journal} {\bibinfo  {journal} {Phys. Rev. A}\ }\textbf {\bibinfo
  {volume} {6}},\ \bibinfo {pages} {1255} (\bibinfo {year} {1972})}\BibitemShut
  {NoStop}%
\bibitem [{\citenamefont {M{\"u}ller}\ \emph {et~al.}(2017)\citenamefont
  {M{\"u}ller}, \citenamefont {Bernhardt}, \citenamefont {Borovik},
  \citenamefont {Buhr}, \citenamefont {Hellhund}, \citenamefont {Holste},
  \citenamefont {Kilcoyne}, \citenamefont {Klumpp}, \citenamefont {Martins},
  \citenamefont {Ricz}, \citenamefont {Seltmann}, \citenamefont {Viefhaus},\
  and\ \citenamefont {Schippers}}]{Muller2017ApJ}%
  \BibitemOpen
  \bibfield  {author} {\bibinfo {author} {\bibfnamefont {A.}~\bibnamefont
  {M{\"u}ller}}, \bibinfo {author} {\bibfnamefont {D.}~\bibnamefont
  {Bernhardt}}, \bibinfo {author} {\bibfnamefont {A.}~\bibnamefont {Borovik}},
  \bibinfo {author} {\bibfnamefont {T.}~\bibnamefont {Buhr}}, \bibinfo {author}
  {\bibfnamefont {J.}~\bibnamefont {Hellhund}}, \bibinfo {author}
  {\bibfnamefont {K.}~\bibnamefont {Holste}}, \bibinfo {author} {\bibfnamefont
  {A.~L.~D.}\ \bibnamefont {Kilcoyne}}, \bibinfo {author} {\bibfnamefont
  {S.}~\bibnamefont {Klumpp}}, \bibinfo {author} {\bibfnamefont
  {M.}~\bibnamefont {Martins}}, \bibinfo {author} {\bibfnamefont
  {S.}~\bibnamefont {Ricz}}, \bibinfo {author} {\bibfnamefont {J.}~\bibnamefont
  {Seltmann}}, \bibinfo {author} {\bibfnamefont {J.}~\bibnamefont {Viefhaus}},\
  and\ \bibinfo {author} {\bibfnamefont {S.}~\bibnamefont {Schippers}},\
  }\bibfield  {title} {\bibinfo {title} {Photoionization of ne atoms and ne$^+$
  ions near the k edge: {PrecisionSpectroscopy} and absolute cross-sections},\
  }\href {https://doi.org/10.3847/1538-4357/836/2/166} {\bibfield  {journal}
  {\bibinfo  {journal} {The Astrophysical Journal}\ }\textbf {\bibinfo {volume}
  {836}},\ \bibinfo {pages} {166} (\bibinfo {year} {2017})}\BibitemShut
  {NoStop}%
\bibitem [{\citenamefont {Sun}\ \emph {et~al.}(2009)\citenamefont {Sun},
  \citenamefont {Liu},\ and\ \citenamefont {Gel'mukhanov}}]{sun2009slowdown}%
  \BibitemOpen
  \bibfield  {author} {\bibinfo {author} {\bibfnamefont {Y.-P.}\ \bibnamefont
  {Sun}}, \bibinfo {author} {\bibfnamefont {J.-C.}\ \bibnamefont {Liu}},\ and\
  \bibinfo {author} {\bibfnamefont {F.}~\bibnamefont {Gel'mukhanov}},\
  }\bibfield  {title} {\bibinfo {title} {Slowdown and compression of a strong
  x-ray free-electron pulse propagating through the mg vapors},\ }\href@noop {}
  {\bibfield  {journal} {\bibinfo  {journal} {EPL (Europhysics Letters)}\
  }\textbf {\bibinfo {volume} {87}},\ \bibinfo {pages} {64002} (\bibinfo {year}
  {2009})}\BibitemShut {NoStop}%
\bibitem [{\citenamefont {Marburger}(1975)}]{marburger1975self}%
  \BibitemOpen
  \bibfield  {author} {\bibinfo {author} {\bibfnamefont {J.}~\bibnamefont
  {Marburger}},\ }\bibfield  {title} {\bibinfo {title} {Self-focusing:
  theory},\ }\href@noop {} {\bibfield  {journal} {\bibinfo  {journal} {Progress
  in quantum electronics}\ }\textbf {\bibinfo {volume} {4}},\ \bibinfo {pages}
  {35} (\bibinfo {year} {1975})}\BibitemShut {NoStop}%
\bibitem [{\citenamefont {Vannucci}\ and\ \citenamefont
  {Teich}(1980)}]{vannucci1980computer}%
  \BibitemOpen
  \bibfield  {author} {\bibinfo {author} {\bibfnamefont {G.}~\bibnamefont
  {Vannucci}}\ and\ \bibinfo {author} {\bibfnamefont {M.~C.}\ \bibnamefont
  {Teich}},\ }\bibfield  {title} {\bibinfo {title} {Computer simulation of
  superposed coherent and chaotic radiation},\ }\href@noop {} {\bibfield
  {journal} {\bibinfo  {journal} {Applied optics}\ }\textbf {\bibinfo {volume}
  {19}},\ \bibinfo {pages} {548} (\bibinfo {year} {1980})}\BibitemShut
  {NoStop}%
\bibitem [{\citenamefont {Weninger}\ and\ \citenamefont
  {Rohringer}(2014)}]{weninger2014transient}%
  \BibitemOpen
  \bibfield  {author} {\bibinfo {author} {\bibfnamefont {C.}~\bibnamefont
  {Weninger}}\ and\ \bibinfo {author} {\bibfnamefont {N.}~\bibnamefont
  {Rohringer}},\ }\bibfield  {title} {\bibinfo {title} {Transient-gain
  photoionization x-ray laser},\ }\href@noop {} {\bibfield  {journal} {\bibinfo
   {journal} {Physical Review A}\ }\textbf {\bibinfo {volume} {90}},\ \bibinfo
  {pages} {063828} (\bibinfo {year} {2014})}\BibitemShut {NoStop}%
\end{thebibliography}%

\end{document}